\DocumentMetadata{uncompress} 
\documentclass[
  10pt,
  twocolumn,
  twoside,
]{article}

\usepackage[numbers,round,comma,sort&compress]{natbib}

\usepackage[
  left=0.65in,
  right=0.65in,
  top=0.65in,
  bottom=0.65in,
]{geometry}

\usepackage{titlesec}
\titleformat{\section}{\raggedright\large\bfseries}{\thesection}{1em}{}
\titleformat{\subsection}{\raggedright\large}{\thesubsection}{1em}{}
\usepackage{ragged2e}

\usepackage[font=small,labelfont=bf]{caption}

\usepackage{tabularx}
\usepackage{longtable}
\usepackage{array}
\usepackage{array}
\newcolumntype{Y}{>{\raggedright\arraybackslash}X}

\usepackage{fancyhdr}
\usepackage[english]{babel}

\usepackage{amssymb}
\usepackage{amsmath}
\usepackage{newunicodechar}
\usepackage{amsfonts}  
\usepackage{array}
\usepackage{nameref}    
\usepackage{hyperref}   
\usepackage{placeins}

\usepackage{verbatim}

\usepackage{environ}

\usepackage{url}
\PassOptionsToPackage{hyphens}{url} 
\makeatletter
\g@addto@macro{\UrlBreaks}{\UrlOrds}
\makeatother

\usepackage[table]{xcolor}   
\usepackage{colortbl}        

\definecolor{goodblue}{RGB}{0, 91, 187}
\hypersetup{
  colorlinks=true,
  allcolors=goodblue,
  urlcolor=goodblue,
  citecolor=goodblue,
  pdfborder={0 0 0},
  breaklinks=true,
}

\usepackage[normalem]{ulem}

\usepackage{textcomp}

\usepackage[export]{adjustbox}

\makeatletter

\def\CT@@do@color{%
  \global\let\CT@do@color\relax
  \@tempdima\wd\z@
  \advance\@tempdima\@tempdimb
  \advance\@tempdima\@tempdimc
  \advance\@tempdimb\tabcolsep
  \advance\@tempdimc\tabcolsep
  \advance\@tempdima2\tabcolsep
  \kern-\@tempdimb
  \leaders\vrule
  \hskip\@tempdima\@plus  1fill
  \kern-\@tempdimc
  \hskip-\wd\z@ \@plus -1fill }
\makeatother

\newcommand{\done}[1]{}

\usepackage{titlesec}
\titleformat*{\paragraph}{\bfseries}

\usepackage{changepage}

\usepackage{graphicx}
\usepackage{epsfig}
\usepackage{verbatim}
\usepackage{enumerate}
\usepackage{enumitem}

\usepackage{ifthen}

\usepackage{longtable}

\usepackage{mathtools}

\usepackage{tabularx}

\newboolean{twocolswitch}

\newcommand{\PreserveBackslash}[1]{\let\temp=\\#1\let\\=\temp}

\newcommand{\sindex}[1]{}
\newcommand{\nindex}[1]{}

\newcommand{\etal}{\textit{et al.}}
\newcommand{\www}[1]{\url{#1}}

\usepackage{lettrine}

\usepackage{changepage}

\usepackage{footmisc}

\usepackage{refcount}

\makeatletter
\newcommand{\footnotemarklabel}[1]{%
  \protect\footnotemark
  \begingroup
    \edef\@currentlabel{\thefootnote}
    \label{#1}%
  \endgroup
}
\makeatother

\renewcommand{\footnoterule}{%
  \kern 3pt
  \hrule width 0.4\columnwidth height 0.4pt
  \kern 6pt
}
\NewEnviron{excerpt}{
    \begin{quote}
      \medskip
      \BODY
      \medskip
    \end{quote}
}

\NewEnviron{editnote}{
    \begin{quote}
      \color{rose}
      $\blacksquare$
      \BODY
      \medskip
    \end{quote}
}

\newcommand{\command}[1]{
  \lstinline[language={[LaTeX]TeX},basicstyle=\ttfamily]{#1}
}

\newcommand{\editbox}[2]{
}

\newcommand{\editboxwithlatex}[2]{
}

\usepackage{fancyvrb}

\usepackage{tcolorbox}
\tcbuselibrary{skins,breakable,listings,breakable}

\usepackage{tikz}
\usetikzlibrary{shapes}

\tikzstyle{mybox} = [draw=lightblue!70, fill=lightblue!7, very thick,
    rectangle, rounded corners, inner sep=10pt, inner ysep=20pt]

\tikzstyle{editortitle} =[draw=archetyperowcoloralt, fill=archetyperowcoloralt, text=black]

\newcommand\Loadedframemethod{default}
\usepackage[framemethod=\Loadedframemethod]{mdframed}

\mdfsetup{skipabove=\topskip,skipbelow=\topskip}

\tikzstyle{loglinetitle} =[draw=icedark, fill=icemedium!50, text=black]

\newenvironment{loglinebox}[1][]{

  \ifstrempty{#1}%
  {\mdfsetup{%
    frametitle={%
       \tikz[baseline=(current bounding box.east),outer sep=0pt]
        \node[loglinetitle, anchor=east,rectangle]
        {\strut~~#1~~\strut};}}
  }%
  {\mdfsetup{%
     frametitle={%
       \tikz[baseline=(current bounding box.east),outer sep=0pt]
        \node[loglinetitle,anchor=east,rectangle]
        {\strut~~#1~~\strut};}}%
   }%
   \mdfsetup{innertopmargin=5pt,linecolor=icedark,%
             linewidth=0.5pt,topline=true,
             frametitleaboveskip=\dimexpr-\ht\strutbox\relax,}
   \begin{mdframed}[backgroundcolor=icelight,nobreak=true]\relax%
     \raggedright
}{\end{mdframed}}

\tikzstyle{abstracttitle} =[draw=magmadark!75, fill=magmamedium!75, text=black]

\newenvironment{abstractbox}[1][]{

  \ifstrempty{#1}%
  {\mdfsetup{%
    frametitle={%
       \tikz[baseline=(current bounding box.east),outer sep=0pt]
        \node[abstracttitle, anchor=east,rectangle]
        {\strut~~#1~~\strut};}}
  }%
  {\mdfsetup{%
     frametitle={%
       \tikz[baseline=(current bounding box.east),outer sep=0pt]
        \node[abstracttitle,anchor=east,rectangle]
        {\strut~~#1~~\strut};}}%
   }%
   \mdfsetup{innertopmargin=5pt,linecolor=magmadark,%
             linewidth=0.5pt,topline=true,
             frametitleaboveskip=\dimexpr-\ht\strutbox\relax,}
   \begin{mdframed}[backgroundcolor=magmalight,nobreak=true]\relax%
     \raggedright
}{\end{mdframed}}

\tikzstyle{infotitle} =[draw=darkgrey, fill=lightgrey!50, text=black]

\newenvironment{infobox}[1][]{

  \ifstrempty{#1}%
  {\mdfsetup{%
    frametitle={%
       \tikz[baseline=(current bounding box.east),outer sep=0pt]
        \node[infotitle, anchor=east,rectangle]
        {\strut~~#1~~\strut};}}
  }%
  {\mdfsetup{%
     frametitle={%
       \tikz[baseline=(current bounding box.east),outer sep=0pt]
        \node[infotitle,anchor=east,rectangle]
        {\strut~~#1~~\strut};}}%
   }%
   \mdfsetup{innertopmargin=5pt,linecolor=grey,%
             linewidth=0.5pt,topline=true,
             frametitleaboveskip=\dimexpr-\ht\strutbox\relax,}
   \begin{mdframed}[backgroundcolor=lightgrey!25,nobreak=true]\relax%
     \raggedright
}{\end{mdframed}}

\tikzstyle{essencetitle} = [draw=magmadark!75, fill=magmamedium!75, text=black]

\usepackage{enumitem}

\tikzstyle{changelogtitle} =[draw=darkgrey, fill=lightgrey!50, text=black]

\usepackage[table]{xcolor}

\definecolor{olivegreen}{rgb}{0.33333,.41961,0.18431}
\definecolor{forestgreen}{rgb}{0.13333,.5451,0.13333}

\definecolor{lightgrey}{rgb}{0.7,0.7,0.7}
\definecolor{verylightgrey}{rgb}{0.90,0.90,0.90}
\definecolor{veryverylightgrey}{rgb}{0.95,0.95,0.95}
\definecolor{grey}{rgb}{0.5,0.5,0.5}
\definecolor{darkgrey}{rgb}{0.3,0.3,0.3}
\definecolor{verydarkgrey}{rgb}{0.15,0.15,0.15}

\definecolor{headerblue}{HTML}{33367E}
\definecolor{unitednationsblue}{HTML}{4D88FF}

\definecolor{charcoal}{HTML}{36454F}
\definecolor{cinerous}{HTML}{98817B}
\definecolor{feldgrau}{HTML}{4D5D53}
\definecolor{glaucous}{HTML}{6082B6}
\definecolor{arsenic}{HTML}{3B444B}
\definecolor{xanadu}{HTML}{738678}

\definecolor{firebrick}{HTML}{B22222}
\definecolor{orangered}{HTML}{FF4500}
\definecolor{tomato}{HTML}{FF6347}

\definecolor{orange}{RGB}{255,116,0}

\definecolor{purpletaupe}{HTML}{3B444B}

\definecolor{rose}{HTML}{E3242B}

\colorlet{editnotecolor}{rose}

\definecolor{headerorange}{RGB}{255,116,0}
\definecolor{headergray}{RGB}{230,230,230}

\definecolor{headerpop}{RGB}{230,230,230}

\definecolor{magmalight}{RGB}{252,251,195}
\definecolor{magmalightalt}{RGB}{250,240,184}
\definecolor{magmamedium}{RGB}{245,200,146}
\definecolor{magmadark}{RGB}{224,106,98}

\definecolor{icelight}{RGB}{223,242,244}
\definecolor{icelightalt}{RGB}{189,222,226}
\definecolor{icemedium}{RGB}{132,184,204}
\definecolor{icedark}{RGB}{103,153,191}

\definecolor{traitrowcolor}{RGB}{223,242,244}
\definecolor{traitrowcoloralt}{RGB}{189,222,226}

\definecolor{characterrowcolor}{RGB}{252,251,195}
\definecolor{characterrowcoloralt}{RGB}{250,240,184}

\definecolor{archetyperowcolor}{RGB}{255,213,212} 
\definecolor{archetyperowcoloralt}{RGB}{255,182,179} 

\definecolor{datasetrowcolor}{RGB}{232,244,234}
\definecolor{datasetrowcoloralt}{RGB}{210,231,214}

\newcommand{\Ncharacters}{2000}
\newcommand{\Ntraits}{464}
\newcommand{\Nstories}{341}

\newcommand{\Ncharactersbase}{2000}

\newcommand{\Ncharactersmain}{2000}
\newcommand{\Ntraitsmain}{464}
\newcommand{\Nstoriesmain}{341}

\newcommand{\Ncharactersmainone}{0800}
\newcommand{\Ntraitsmainone}{235}
\newcommand{\Nstoriesmainone}{090}

\newcommand{\Ncharactersmaintwo}{1600}
\newcommand{\Ntraitsmaintwo}{364}
\newcommand{\Nstoriesmaintwo}{241}

\newcommand{\onlinesiteshort}{compstorylab.org/archetypometrics}
\newcommand{\onlinesite}{https://\onlinesiteshort}

\newcommand{\semdiffsign}{\Leftrightarrow}
\newcommand{\semdiffsignleft}{\Leftarrow}
\newcommand{\semdiffsignright}{\Rightarrow}

\newcommand{\semdiff}[2]{\{#1\,$\semdiffsign$\,#2\}}

\newcommand{\semdiffbold}[2]{\{\textbf{#1}\,$\semdiffsign$\,\textbf{#2}\}}
\newcommand{\semdiffboldleft}[2]{\{\textbf{#1}\,$\semdiffsignleft$\,#2\}}
\newcommand{\semdiffboldright}[2]{\{#1\,$\semdiffsignright$\,\textbf{#2}\}}

\newcommand{\semdiffmath}[2]{\{\textnormal{\textbf{#1}}\!\semdiffsign\!{\textnormal{\textbf{#2}}\}}}
\newcommand{\semdiffmathleft}[2]{\{\textnormal{\textbf{#1}}\!\semdiffsignleft\!{\textnormal{#2}\}}}
\newcommand{\semdiffmathright}[2]{\{\textnormal{#1}\!\semdiffsignright\!{\textnormal{\textbf{#2}}\}}}

\newcommand{\datasetsymbol}{\colorbox{datasetrowcolor}{\textcolor{black}{\stackanchor{\scalebox{\babaisyouboxscale}{DA}}{\scalebox{\babaisyouboxscale}{TA}}}}}

\newcommand{\dataset}[1]{\mbox{\datasetsymbol}_{#1}}

\newcommand{\datasetbase}[1]{
    \IfEqCase{#1}{
        {0800}{\datasetsymbol{1}}
        {1600}{\datasetsymbol{2}}
        {2000}{\datasetsymbol{3}}
    }[\PackageError{datasetbase}{Undefined option to datasetbase: #1}{}]%
}%

\newcommand{\datasetNcharacters}[1]{
    \IfEqCase{#1}{
        {1}{800}
        {2}{1600}
        {3}{2000}
    }[\PackageError{datasetNcharacters}{Undefined option to datasetNcharacters: #1}{}]%
}%

\newcommand{\datasetNtraits}[1]{
    \IfEqCase{#1}{
        {1}{235}
        {2}{364}
        {3}{464}
    }[\PackageError{datasetNtraits}{Undefined option to datasetNtraits: #1}{}]%
}%

\newcommand{\datasetNstories}[1]{
    \IfEqCase{#1}{
        {1}{90}
        {2}{241}
        {3}{341}
    }[\PackageError{datasetNstories}{Undefined option to datasetNstories: #1}{}]%
}%

\newcommand{\padzero}[1]{\ifnum #1 < 10 0\fi #1}

\usepackage{siunitx}
\usepackage{multirow}
\usepackage{booktabs} 

\newcommand\zeropad[2]{%
  \ifnum#2<0\relax%
    {\ensuremath-}\zeropadA{#1}{\the\numexpr#2*-1\relax}%
  \else%
    \zeropadA{#1}{#2}%
  \fi%
}
\def\zeropadA#1#2{%
  \ifnum1#2<1#1
    \zeropadA{#1}{0#2}%
  \else%
    #2%
  \fi%
}

\usepackage{xstring}

\newcommand{\archetypesemdiff}[1]{
    \IfEqCase{#1}{
        {1}{\semdiffbold{\archetypelinkbase{Fool}}{\archetypelinkbase{Hero}}}                  
        {2}{\semdiffbold{\archetypelinkbase{Angel}}{\archetypelinkbase{Demon}}}                
        {3}{\semdiffbold{\archetypelinkbase{Traditionalist}}{\archetypelinkbase{Adventurer}}}  
        {4}{\semdiffbold{\archetypelinksimple{Lone-Wolf}{Lone~Wolf}}{\archetypelinkbase{Diva}}}
        {5}{\semdiffbold{\archetypelinkbase{Outcast}}{\archetypelinkbase{Sophisticate}}}       
        {6}{\semdiffbold{\archetypelinkbase{Brute}}{\archetypelinkbase{Geek}}}                 
    }[\PackageError{archetypesemdiff}{Undefined option to archetypesemdiff: #1}{}]%
}%

\newcommand{\archetypesemdiffleft}[1]{
    \IfEqCase{#1}{
        {1}{\semdiffboldleft{\archetypelinkbase{Fool}}{\archetypelinkbase{Hero}}}                  
        {2}{\semdiffboldleft{\archetypelinkbase{Angel}}{\archetypelinkbase{Demon}}}                
        {3}{\semdiffboldleft{\archetypelinkbase{Traditionalist}}{\archetypelinkbase{Adventurer}}}  
        {4}{\semdiffboldleft{\archetypelinksimple{Lone-Wolf}{Lone~Wolf}}{\archetypelinkbase{Diva}}}
        {5}{\semdiffboldleft{\archetypelinkbase{Outcast}}{\archetypelinkbase{Sophisticate}}}       
        {6}{\semdiffboldleft{\archetypelinkbase{Brute}}{\archetypelinkbase{Geek}}}                 
    }[\PackageError{archetypesemdiff}{Undefined option to archetypesemdiff: #1}{}]%
}%

\newcommand{\archetypesemdiffright}[1]{
    \IfEqCase{#1}{
        {1}{\semdiffboldright{\archetypelinkbase{Fool}}{\archetypelinkbase{Hero}}}                  
        {2}{\semdiffboldright{\archetypelinkbase{Angel}}{\archetypelinkbase{Demon}}}                
        {3}{\semdiffboldright{\archetypelinkbase{Traditionalist}}{\archetypelinkbase{Adventurer}}}  
        {4}{\semdiffboldright{\archetypelinksimple{Lone-Wolf}{Lone~Wolf}}{\archetypelinkbase{Diva}}}
        {5}{\semdiffboldright{\archetypelinkbase{Outcast}}{\archetypelinkbase{Sophisticate}}}       
        {6}{\semdiffboldright{\archetypelinkbase{Brute}}{\archetypelinkbase{Geek}}}                 
    }[\PackageError{archetypesemdiff}{Undefined option to archetypesemdiff: #1}{}]%
}%

\newcommand{\archetypesemdiffmath}[1]{
  \IfEqCase{#1}{
    {1}{\semdiffmath{\archetypelinkbase{Fool}}{\archetypelinkbase{Hero}}}
    {2}{\semdiffmath{\archetypelinkbase{Angel}}{\archetypelinkbase{Demon}}}
    {3}{\semdiffmath{\archetypelinkbase{Traditionalist}}{\archetypelinkbase{Adventurer}}}
    {4}{\semdiffmath{\archetypelinksimple{Lone-Wolf}{Lone~Wolf}}{\archetypelinkbase{Diva}}}
    {5}{\semdiffmath{\archetypelinkbase{Outcast}}{\archetypelinkbase{Sophisticate}}}
    {6}{\semdiffmath{\archetypelinkbase{Brute}}{\archetypelinkbase{Geek}}}                 
  }[\PackageError{archetypesemdiffmath}{Undefined option to archetypesemdiffmath: #1}{}]%
}%

\newcommand{\archetypesemdiffmathswap}[1]{
  \IfEqCase{#1}{
    {1}{\semdiffmath{\archetypelinkbase{Hero}}{\archetypelinkbase{Fool}}}
    {2}{\semdiffmath{\archetypelinkbase{Demon}}{\archetypelinkbase{Angel}}}
    {3}{\semdiffmath{\archetypelinkbase{Adventurer}}{\archetypelinkbase{Traditionalist}}}
    {4}{\semdiffmath{\archetypelinkbase{Diva}}{\archetypelinksimple{Lone-Wolf}{Lone~Wolf}}}
    {5}{\semdiffmath{\archetypelinkbase{Sophisticate}}{\archetypelinkbase{Outcast}}}
    {6}{\semdiffmath{{\archetypelinkbase{Geek}}\archetypelinkbase{Brute}}}
  }[\PackageError{archetypesemdiffmathswap}{Undefined option to archetypesemdiffmathswap: #1}{}]%
}%

\newcommand{\archetypesemdiffmathleft}[1]{
  \IfEqCase{#1}{
    {1}{\semdiffmathleft{\archetypelinkbase{Fool}}{\archetypelinkbase{Hero}}}                  
    {2}{\semdiffmathleft{\archetypelinkbase{Angel}}{\archetypelinkbase{Demon}}}                
    {3}{\semdiffmathleft{\archetypelinkbase{Traditionalist}}{\archetypelinkbase{Adventurer}}}  
    {4}{\semdiffmathleft{\archetypelinksimple{Lone-Wolf}{Lone~Wolf}}{\archetypelinkbase{Diva}}}
    {5}{\semdiffmathleft{\archetypelinkbase{Outcast}}{\archetypelinkbase{Sophisticate}}}       
    {6}{\semdiffmathleft{\archetypelinkbase{Brute}}{\archetypelinkbase{Geek}}}                 
  }[\PackageError{archetypesemdiffmathleft}{Undefined option to archetypesemdiffmathleft: #1}{}]%
}%

\newcommand{\archetypesemdiffmathright}[1]{
  \IfEqCase{#1}{
    {1}{\semdiffmathright{\archetypelinkbase{Fool}}{\archetypelinkbase{Hero}}}                  
    {2}{\semdiffmathright{\archetypelinkbase{Angel}}{\archetypelinkbase{Demon}}}                
    {3}{\semdiffmathright{\archetypelinkbase{Traditionalist}}{\archetypelinkbase{Adventurer}}}  
    {4}{\semdiffmathright{\archetypelinksimple{Lone-Wolf}{Lone~Wolf}}{\archetypelinkbase{Diva}}}
    {5}{\semdiffmathright{\archetypelinkbase{Outcast}}{\archetypelinkbase{Sophisticate}}}       
    {6}{\semdiffmathright{\archetypelinkbase{Brute}}{\archetypelinkbase{Geek}}}                 
  }[\PackageError{archetypesemdiffmathright}{Undefined option to archetypesemdiffmathright: #1}{}]%
}%

\newcommand{\essentialsemdiff}[1]{
  \IfEqCase{#1}{
    {1}{\semdiff{\essentialtraitlinknegative{1}{weak/incompetent/lazy/stupid}}{\essentialtraitlinkpositive{1}{powerful/capable/purposeful/intelligent}}}
    {2}{\semdiff{\essentialtraitlinknegative{2}{safe/pure/virtuous/humble}}{\essentialtraitlinkpositive{2}{dangerous/depraved/corrupt/arrogant}}}
    {3}{\semdiff{\essentialtraitlinknegative{3}{serious/predictable/humorless/uncreative}}{\essentialtraitlinkpositive{3}{playful/unpredictable/funny/creative}}}
    {4}{\semdiff{\essentialtraitlinknegative{4}{rugged/stoic/independent/blunt}}{\essentialtraitlinkpositive{4}{refined/dramatic/dependent/sensitive}}}
    {5}{\semdiff{\essentialtraitlinknegative{5}{unlucky/unsophisticated/traumatized}}{\essentialtraitlinkpositive{5}{fortunate/sophisticated/confident}}}
    {6}{\semdiff{\essentialtraitlinknegative{6}{physical/mainstream/simple-minded}}{\essentialtraitlinkpositive{6}{intellectual/weird/complex}}}
    {7}{\semdiff{\essentialtraitlinknegative{7}{dramatic/attractive/young}}{\essentialtraitlinkpositive{7}{comedic/ugly/old}}}
    {8}{\semdiff{\essentialtraitlinknegative{8}{spiritual/rural/historical}}{\essentialtraitlinkpositive{8}{skeptical/urban/modern}}}
    {9}{\semdiff{\essentialtraitlinknegative{9}{old/historical/low-tempo}}{\essentialtraitlinkpositive{9}{young/modern/high-tempo}}}
    {10}{\semdiff{\essentialtraitlinknegative{10}{feminine/luddite}}{\essentialtraitlinkpositive{10}{masculine/technophile}}}
    {11}{\semdiff{\essentialtraitlinknegative{11}{secondary/street-wise}}{\essentialtraitlinkpositive{11}{primary/sheltered}}}
  }[\PackageError{essentialsemdiff}{Undefined option to essentialsemdiff: #1}{}]%
}%

\newcommand{\essentialsemdiffloose}[1]{
  \IfEqCase{#1}{
    {1}{\semdiff{\essentialtraitlinknegative{1}{weak, incompetent, lazy, stupid}}{\essentialtraitlinkpositive{1}{powerful, capable, purposeful, intelligent}}}
    {2}{\semdiff{\essentialtraitlinknegative{2}{safe, pure, virtuous, humble}}{\essentialtraitlinkpositive{2}{dangerous, depraved, corrupt, arrogant}}}
    {3}{\semdiff{\essentialtraitlinknegative{3}{serious, predictable, humorless, uncreative}}{\essentialtraitlinkpositive{3}{playful, unpredictable, funny, creative}}}
    {4}{\semdiff{\essentialtraitlinknegative{4}{rugged, stoic, independent, blunt}}{\essentialtraitlinkpositive{4}{refined, dramatic, dependent, sensitive}}}
    {5}{\semdiff{\essentialtraitlinknegative{5}{unlucky, unsophisticated, traumatized}}{\essentialtraitlinkpositive{5}{fortunate, sophisticated, confident}}}
    {6}{\semdiff{\essentialtraitlinknegative{6}{physical, mainstream, simple-minded}}{\essentialtraitlinkpositive{6}{intellectual, weird, complex}}}
    {7}{\semdiff{\essentialtraitlinknegative{7}{dramatic, attractive, young}}{\essentialtraitlinkpositive{7}{comedic, ugly, old}}}
    {8}{\semdiff{\essentialtraitlinknegative{8}{spiritual, rural, historical}}{\essentialtraitlinkpositive{8}{skeptical, urban, modern}}}
    {9}{\semdiff{\essentialtraitlinknegative{9}{old, historical, low-tempo}}{\essentialtraitlinkpositive{9}{young, modern, high-tempo}}}
    {10}{\semdiff{\essentialtraitlinknegative{10}{feminine, luddite}}{\essentialtraitlinkpositive{10}{masculine, technophile}}}
    {11}{\semdiff{\essentialtraitlinknegative{11}{secondary, street-wise}}{\essentialtraitlinkpositive{11}{primary, sheltered}}}
  }[\PackageError{essentialsemdiffloose}{Undefined option to essentialsemdiffloose: #1}{}]%
}%

\newcommand{\essentialsemdifflooseleft}[1]{
    \IfEqCase{#1}{
        {1}{\semdiff{\textbf{\essentialtraitlinknegative{1}{weak, incompetent, lazy, stupid}}}{\essentialtraitlinkpositive{1}{powerful, capable, purposeful, intelligent}}}
        {2}{\semdiff{\textbf{\essentialtraitlinknegative{2}{safe, pure, virtuous, humble}}}{\essentialtraitlinkpositive{2}{dangerous, depraved, corrupt, arrogant}}}
        {3}{\semdiff{\textbf{\essentialtraitlinknegative{3}{serious, predictable, humorless, uncreative}}}{\essentialtraitlinkpositive{3}{playful, unpredictable, funny, creative}}}
        {4}{\semdiff{\textbf{\essentialtraitlinknegative{4}{rugged, stoic, independent, blunt}}}{\essentialtraitlinkpositive{4}{refined, dramatic, dependent, sensitive}}}
        {5}{\semdiff{\textbf{\essentialtraitlinknegative{5}{unlucky, unsophisticated, traumatized}}}{\essentialtraitlinkpositive{5}{fortunate, sophisticated, confident}}}
        {6}{\semdiff{\textbf{\essentialtraitlinknegative{6}{physical, mainstream, simple-minded}}}{\essentialtraitlinkpositive{6}{intellectual, weird, complex}}}
        {7}{\semdiff{\textbf{\essentialtraitlinknegative{7}{dramatic, attractive, young}}}{\essentialtraitlinkpositive{7}{comedic, ugly, old}}}
        {8}{\semdiff{\textbf{\essentialtraitlinknegative{8}{spiritual, rural, historical}}}{\essentialtraitlinkpositive{8}{skeptical, urban, modern}}}
        {9}{\semdiff{\textbf{\essentialtraitlinknegative{9}{old, historical, low-tempo}}}{\essentialtraitlinkpositive{9}{young, modern, high-tempo}}}
        {10}{\semdiff{\textbf{\essentialtraitlinknegative{10}{feminine, luddite}}}{\essentialtraitlinkpositive{10}{masculine, technophile}}}
        {11}{\semdiff{\textbf{\essentialtraitlinknegative{11}{secondary, street-wise}}}{\essentialtraitlinkpositive{11}{primary, sheltered}}}
    }[\PackageError{essentialsemdifflooseleft}{Undefined option to essentialsemdifflooseleft: #1}{}]%
 }%

\newcommand{\essentialsemdifflooseright}[1]{
  \IfEqCase{#1}{
    {1}{\semdiff{\essentialtraitlinknegative{1}{weak, incompetent, lazy, stupid}}{\textbf{\essentialtraitlinkpositive{1}{powerful, capable, purposeful, intelligent}}}}
    {2}{\semdiff{\essentialtraitlinknegative{2}{safe, pure, virtuous, humble}}{\textbf{\essentialtraitlinkpositive{2}{dangerous, depraved, corrupt, arrogant}}}}
    {3}{\semdiff{\essentialtraitlinknegative{3}{serious, predictable, humorless, uncreative}}{\textbf{\essentialtraitlinkpositive{3}{playful, unpredictable, funny, creative}}}}
    {4}{\semdiff{\essentialtraitlinknegative{4}{rugged, stoic, independent, blunt}}{\textbf{\essentialtraitlinkpositive{4}{refined, dramatic, dependent, sensitive}}}}
    {5}{\semdiff{\essentialtraitlinknegative{5}{unlucky, unsophisticated, traumatized}}{\textbf{\essentialtraitlinkpositive{5}{fortunate, sophisticated, confident}}}}
    {6}{\semdiff{\essentialtraitlinknegative{6}{physical, mainstream, simple-minded}}{\textbf{\essentialtraitlinkpositive{6}{intellectual, weird, complex}}}}
    {7}{\semdiff{\essentialtraitlinknegative{7}{dramatic, attractive, young}}{\textbf{\essentialtraitlinkpositive{7}{comedic, ugly, old}}}}
    {8}{\semdiff{\essentialtraitlinknegative{8}{spiritual, historical, rural}}{\textbf{\essentialtraitlinkpositive{8}{skeptical, urban, modern}}}}
    {9}{\semdiff{\essentialtraitlinknegative{9}{old, historical, low-tempo}}{\textbf{\essentialtraitlinkpositive{9}{young, modern, high-tempo}}}}
    {10}{\semdiff{\essentialtraitlinknegative{10}{feminine, luddite}}{\textbf{\essentialtraitlinkpositive{10}{masculine, technophile}}}}
    {11}{\semdiff{\essentialtraitlinknegative{11}{secondary, street-wise}}{\textbf{\essentialtraitlinkpositive{11}{primary, sheltered}}}}
  }[\PackageError{essentialsemdifflooseright}{Undefined option to essentialsemdifflooseright: #1}{}]%
}%

\newcommand{\essentialsemdiffmathleft}[1]{
    \IfEqCase{#1}{
        {1}{\semdiffmathleft{weak/incompetent/lazy/stupid}{powerful/capable/purposeful/intelligent}}
        {2}{\semdiffmathleft{safe/pure/virtuous/humble}{dangerous/depraved/corrupt/arrogant}}
        {3}{\semdiffmathleft{serious/predictable/humorless/uncreative}{playful/unpredictable/funny/creative}}
        {4}{\semdiffmathleft{rugged/stoic/independent/blunt}{refined/dramatic/dependent/sensitive}}
        {5}{\semdiffmathleft{unlucky/unsophisticated/traumatized}{fortunate/sophisticated/confident}}
        {6}{\semdiffmathleft{physical/mainstream/simple-minded}{intellectual/weird/complex}}
        {7}{\semdiffmathleft{dramatic/attractive/young}{comedic/ugly/old}}
        {8}{\semdiffmathleft{spiritual/rural/historical}{skeptical/urban/modern}}
        {9}{\semdiffmathleft{old/historical/low-tempo}{young/modern/high-tempo}}
        {10}{\semdiffmathleft{feminine/luddite}{masculine/technophile}}
        {11}{\semdiffmathleft{secondary/street-wise}{primary/sheltered}}
    }[\PackageError{essentialsemdiffmathleft}{Undefined option to essentialsemdiffmathleft: #1}{}]%
}%

\newcommand{\essentialsemdiffmathright}[1]{
    \IfEqCase{#1}{
        {1}{\semdiffmathright{weak/incompetent/lazy/stupid}{powerful/capable/purposeful/intelligent}}
        {2}{\semdiffmathright{safe/pure/virtuous/humble}{dangerous/depraved/corrupt/arrogant}}
        {3}{\semdiffmathright{serious/predictable/humorless/uncreative}{playful/unpredictable/funny/creative}}
        {4}{\semdiffmathright{rugged/stoic/independent/blunt}{refined/dramatic/dependent/sensitive}}
        {5}{\semdiffmathright{unlucky/unsophisticated/traumatized}{fortunate/sophisticated/confident}}
        {6}{\semdiffmathright{physical/mainstream/simple-minded}{intellectual/weird/complex}}
        {7}{\semdiffmathright{dramatic/attractive/young}{comedic/ugly/old}}
        {8}{\semdiffmathright{spiritual/rural/historical}{skeptical/urban/modern}}
        {9}{\semdiffmathright{old/historical/low-tempo}{young/modern/high-tempo}}
        {10}{\semdiffmathright{feminine/luddite}{masculine/technophile}}
        {11}{\semdiffmathright{secondary/street-wise}{primary/sheltered}}
    }[\PackageError{essentialsemdiffmathright}{Undefined option to essentialsemdiffmathright: #1}{}]%
}%

\newcommand{\essentialsemdiffmath}[1]{
    \IfEqCase{#1}{
        {1}{\semdiffmath{weak/incompetent/lazy/stupid}{powerful/capable/purposeful/intelligent}}
        {2}{\semdiffmath{safe/pure/virtuous/humble}{dangerous/depraved/corrupt/arrogant}}
        {3}{\semdiffmath{serious/predictable/humorless/uncreative}{playful/unpredictable/funny/creative}}
        {4}{\semdiffmath{rugged/stoic/independent/blunt}{refined/dramatic/dependent/sensitive}}
        {5}{\semdiffmath{unlucky/unsophisticated/traumatized}{fortunate/sophisticated/confident}}
        {6}{\semdiffmath{physical/mainstream/simple-minded}{intellectual/weird/complex}}
        {7}{\semdiffmath{dramatic/attractive/young}{comedic/ugly/old}}
        {8}{\semdiffmath{spiritual/rural/historical}{skeptical/urban/modern}}
        {9}{\semdiffmath{old/historical/low-tempo}{young/modern/high-tempo}}
        {10}{\semdiffmath{feminine/luddite}{masculine/technophile}}
        {11}{\semdiffmath{secondary/street-wise}{primary/sheltered}}
    }[\PackageError{essentialsemdiffmath}{Undefined option to essentialsemdiffmath: #1}{}]%
}%

\newcommand{\ousiometricsemdiff}[1]{
    \IfEqCase{#1}{
        {1}{\semdiffbold{weak}{powerful}}
        {2}{\semdiffbold{safe}{dangerous}}
        {3}{\semdiffbold{structured}{unstructured}}
    }[\PackageError{ousiometricsemdiff}{Undefined option to ousiometricsemdiff: #1}{}]%
}%

\newcommand{\ousiometricsemdiffmath}[1]{
    \IfEqCase{#1}{
        {1}{\semdiffmath{weak}{powerful}}
        {2}{\semdiffmath{safe}{dangerous}}
        {3}{\semdiffmath{structured}{unstructured}}
    }[\PackageError{ousiometricsemdiffmath}{Undefined option to ousiometricsemdiffmath: #1}{}]%
}%

\newcommand{\ousiometricsemdiffmathleft}[1]{
    \IfEqCase{#1}{
        {1}{\semdiffmathleft{weak}{powerful}}
        {2}{\semdiffmathleft{safe}{dangerous}}
        {3}{\semdiffmathleft{structured}{unstructured}}
    }[\PackageError{ousiometricsemdiffmathleft}{Undefined option to ousiometricsemdiffmathleft: #1}{}]%
}%

\newcommand{\ousiometricsemdiffmathright}[1]{
    \IfEqCase{#1}{
        {1}{\semdiffmathright{weak}{powerful}}
        {2}{\semdiffmathright{safe}{dangerous}}
        {3}{\semdiffmathright{structured}{unstructured}}
    }[\PackageError{ousiometricsemdiffmathright}{Undefined option to ousiometricsemdiffmathright: #1}{}]%
}%

\newcommand{\dimensiontype}[1]{
    \IfEqCase{#1}{
        {1}{Primary archetype}
        {2}{Primary archetype}
        {3}{Primary archetype}
        {4}{Secondary Archetype}
        {5}{Secondary Archetype}
        {6}{Secondary Archetype}
        {7}{Complex Trait}
        {8}{Complex Trait}
        {9}{Complex Trait}
        {10}{Complex Trait}
        {11}{Complex Trait}
    }[\PackageError{dimensiontype}{Undefined option to dimensiontype: #1}{}]%
}%

\usepackage{stackengine}
\setstackgap{S}{1pt}
\newcommand{\babaisyouboxscale}{0.48}

\newcommand{\archetyperatiosymbol}[2]{{\colorbox{#1}{\textcolor{#2}{\stackanchor{\scalebox{\babaisyouboxscale}{AR}}{\scalebox{\babaisyouboxscale}{CH}}}}}}

\newcommand{\appendixsymbol}[2]{{\colorbox{#1}{\textcolor{#2}{\stackanchor{\scalebox{\babaisyouboxscale}{AP}}{\scalebox{\babaisyouboxscale}{DX}}}}}}

\newcommand{\archetypometricshome}{\onlinesite}
\newcommand{\cardsdir}{\archetypometricshome/cards}

\newcommand{\onlinelinksymbol}{\nnearrow}
\newcommand{\externallinksymbol}{{\tiny$^{{}_{\onlinelinksymbol}}$\!\!}}
\newcommand{\internallinksymbol}{{$^{\Rsh}$\!\!}}
\newcommand{\paperlinksymbol}{\externallinksymbol}

\usepackage{xspace}

\usepackage{stmaryrd}

\newcommand{\characterlinksimpledataset}[3]{
  \IfEqCase{#3}{
    {1}{\href{\cardsdir/#1-\Ncharactersmainone-\Ntraitsmainone-\Nstoriesmainone.pdf}{\textcolor{verydarkgrey}{#2\colorbox{datasetrowcolor}{$\dataset{#3}$}\paperlinksymbol}}}
    {2}{\href{\cardsdir/#1-\Ncharactersmaintwo-\Ntraitsmaintwo-\Nstoriesmaintwo.pdf}{\textcolor{verydarkgrey}{#2\colorbox{datasetrowcolor}{$\dataset{#3}$}\paperlinksymbol}}}
    {3}{\href{\cardsdir/#1-\Ncharactersmain-\Ntraitsmain-\Nstoriesmain.pdf}{\textcolor{verydarkgrey}{#2\colorbox{datasetrowcolor}{$\dataset{#3}$}\paperlinksymbol}}}
  }[\PackageError{characterlinksimpledataset}{Undefined option to characterlinksimpledataset: #1}{}]%
}

\newcommand{\traitlinksimpledataset}[3]{
  \IfEqCase{#3}{
    {1}{\textcolor{verydarkgrey}{\semdiff{\href{\cardsdir/#2--#1-\Ncharactersmainone-\Ntraitsmainone-\Nstoriesmainone.pdf}{#1\colorbox{datasetrowcolor}{$\dataset{#3}$}\paperlinksymbol}}{\href{\cardsdir/#1--#2-\Ncharactersmainone-\Ntraitsmainone-\Nstoriesmainone.pdf}{#2\colorbox{datasetrowcolor}{$\dataset{#3}$}\paperlinksymbol}}}}
    {2}{\textcolor{verydarkgrey}{\semdiff{\href{\cardsdir/#2--#1-\Ncharactersmaintwo-\Ntraitsmaintwo-\Nstoriesmaintwo.pdf}{#1\colorbox{datasetrowcolor}{$\dataset{#3}$}\paperlinksymbol}}{\href{\cardsdir/#1--#2-\Ncharactersmaintwo-\Ntraitsmaintwo-\Nstoriesmaintwo.pdf}{#2\colorbox{datasetrowcolor}{$\dataset{#3}$}\paperlinksymbol}}}}
    {3}{\textcolor{verydarkgrey}{\semdiff{\href{\cardsdir/#2--#1-\Ncharactersmain-\Ntraitsmain-\Nstoriesmain.pdf}{#1\colorbox{datasetrowcolor}{$\dataset{#3}$}\paperlinksymbol}}{\href{\cardsdir/#1--#2-\Ncharactersmain-\Ntraitsmain-\Nstoriesmain.pdf}{#2\colorbox{datasetrowcolor}{$\dataset{#3}$}\paperlinksymbol}}}}
  }[\PackageError{traitlinksimpledataset}{Undefined option to traitlinksimpledataset: #1}{}]%
}

\newcommand{\traitlinksimpledatasetalt}[5]{
  \IfEqCase{#5}{
    {1}{\textcolor{verydarkgrey}{\semdiff{\href{\cardsdir/#2--#1-\Ncharactersmainone-\Ntraitsmainone-\Nstoriesmainone.pdf}{#3\colorbox{datasetrowcolor}{$\dataset{#5}$}\paperlinksymbol}}{\href{\cardsdir/#1--#2-\Ncharactersmainone-\Ntraitsmainone-\Nstoriesmainone.pdf}{#4\colorbox{datasetrowcolor}{$\dataset{#5}$}\paperlinksymbol}}}}
    {2}{\textcolor{verydarkgrey}{\semdiff{\href{\cardsdir/#2--#1-\Ncharactersmaintwo-\Ntraitsmaintwo-\Nstoriesmaintwo.pdf}{#3\colorbox{datasetrowcolor}{$\dataset{#5}$}\paperlinksymbol}}{\href{\cardsdir/#1--#2-\Ncharactersmaintwo-\Ntraitsmaintwo-\Nstoriesmaintwo.pdf}{#4\colorbox{datasetrowcolor}{$\dataset{#5}$}\paperlinksymbol}}}}
    {3}{\textcolor{verydarkgrey}{\semdiff{\href{\cardsdir/#2--#1-\Ncharactersmain-\Ntraitsmain-\Nstoriesmain.pdf}{#3\colorbox{datasetrowcolor}{$\dataset{#5}$}\paperlinksymbol}}{\href{\cardsdir/#1--#2-\Ncharactersmain-\Ntraitsmain-\Nstoriesmain.pdf}{#4\colorbox{datasetrowcolor}{$\dataset{#5}$}\paperlinksymbol}}}}
  }[\PackageError{traitlinksimpledatasetalt}{Undefined option to traitlinksimpledatasetalt: #1}{}]%
}

\newcommand{\storylinksimpledataset}[3]{
  \IfEqCase{#3}{
    {1}{\href{\cardsdir/#1-\Ncharactersmainone-\Ntraitsmainone-\Nstoriesmainone.pdf}{\textcolor{verydarkgrey}{#2\colorbox{datasetrowcolor}{$\dataset{#3}$}\paperlinksymbol}}}
    {2}{\href{\cardsdir/#1-\Ncharactersmaintwo-\Ntraitsmaintwo-\Nstoriesmaintwo.pdf}{\textcolor{verydarkgrey}{#2\colorbox{datasetrowcolor}{$\dataset{#3}$}\paperlinksymbol}}}
    {3}{\href{\cardsdir/#1-\Ncharactersmain-\Ntraitsmain-\Nstoriesmain.pdf}{\textcolor{verydarkgrey}{#2\colorbox{datasetrowcolor}{$\dataset{#3}$}\paperlinksymbol}}}
  }[\PackageError{storylinksimpledataset}{Undefined option to storylinksimpledataset: #1}{}]%
}

\newcommand{\grouplinksimpledataset}[3]{
  \IfEqCase{#3}{
    {1}{\href{\cardsdir/#1-\Ncharactersmainone-\Ntraitsmainone-\Nstoriesmainone.pdf}{\textcolor{verydarkgrey}{#2\colorbox{datasetrowcolor}{$\dataset{#3}$}\paperlinksymbol}}}
    {2}{\href{\cardsdir/#1-\Ncharactersmaintwo-\Ntraitsmaintwo-\Nstoriesmaintwo.pdf}{\textcolor{verydarkgrey}{#2\colorbox{datasetrowcolor}{$\dataset{#3}$}\paperlinksymbol}}}
    {3}{\href{\cardsdir/#1-\Ncharactersmain-\Ntraitsmain-\Nstoriesmain.pdf}{\textcolor{verydarkgrey}{#2\colorbox{datasetrowcolor}{$\dataset{#3}$}\paperlinksymbol}}}
  }[\PackageError{grouplinksimpledataset}{Undefined option to grouplinksimpledataset: #1}{}]%
}

\newcommand{\archetypelinkbase}[1]{\href{\cardsdir/Archetype-#1-component-size-\Ncharacters-\Ntraits-\Nstories.pdf}{\textcolor{verydarkgrey}{#1\paperlinksymbol}}}

\newcommand{\archetypelinksimple}[2]{\href{\cardsdir/Archetype-#1-component-size-\Ncharacters-\Ntraits-\Nstories.pdf}{\textcolor{verydarkgrey}{#2\paperlinksymbol}}}

\newcommand{\archetypelinksimpledataset}[3]{
  \IfEqCase{#3}{
    {1}{\href{\cardsdir/Archetype-#1-component-size-\Ncharactersmainone-\Ntraitsmainone-\Nstoriesmainone.pdf}{\textcolor{verydarkgrey}{#2\colorbox{datasetrowcolor}{$\dataset{#3}$}\paperlinksymbol}}}
    {2}{\href{\cardsdir/Archetype-#1-component-size-\Ncharactersmaintwo-\Ntraitsmaintwo-\Nstoriesmaintwo.pdf}{\textcolor{verydarkgrey}{#2\colorbox{datasetrowcolor}{$\dataset{#3}$}\paperlinksymbol}}}
    {3}{\href{\cardsdir/Archetype-#1-component-size-\Ncharactersmain-\Ntraitsmain-\Nstoriesmain.pdf}{\textcolor{verydarkgrey}{#2\colorbox{datasetrowcolor}{$\dataset{#3}$}\paperlinksymbol}}}
  }[\PackageError{archetypelinksimpledataset}{Undefined option to archetypelinksimpledataset: #1}{}]%
}

\newcommand{\archetypelinkratiosimpleappendix}[2]{{\hypersetup{allcolors=.}\hyperref[page:N\Ncharactersbase_archetypometrics.archetypeclass-#1]{#2\archetyperatiosymbol{archetyperowcolor}{black}\,\appendixsymbol{verydarkgrey}{white}\internallinksymbol}}}

\newcommand{\essentialtraitlinknegative}[2]{\href{\cardsdir/Essential-Trait-\zeropad{000}{#1}-negative-component-size-\Ncharacters-\Ntraits-\Nstories.pdf}{\textcolor{verydarkgrey}{#2\paperlinksymbol}}}

\newcommand{\essentialtraitlinkpositive}[2]{\href{\cardsdir/Essential-Trait-\zeropad{000}{#1}-positive-component-size-\Ncharacters-\Ntraits-\Nstories.pdf}{\textcolor{verydarkgrey}{#2\paperlinksymbol}}}

\newcommand{\characterlink}[1]{
  \IfEqCase{#1}{
    }[\PackageError{characterlink}{Undefined option to characterlink: #1}{}]%
}%

\newcommand{\characterlinkinsert}[2]{
  \IfEqCase{#1}{
    }[\PackageError{characterlinkinsert}{Undefined option to characterlinkinsert: #1}{}]%
}%

\setboolean{twocolswitch}{true}

\usepackage{natbib}
\setcitestyle{square}
\raggedright

\usepackage{pdfpages}
\usepackage{newpax}
\newpaxsetup{usefileattributes=true}

\usepackage{authblk}

\begin{document}

\title{\protect
Interaction dynamics in `Friends’
}
\onecolumn


\renewcommand*{\Authsep}{, }
\renewcommand*{\Authand}{, }
\renewcommand*{\Authands}{, }
\renewcommand*{\Affilfont}{\normalsize\normalfont}
\renewcommand*{\Authfont}{\bfseries}
\setlength{\affilsep}{2em}

\renewcommand*{\Authsep}{, }
\renewcommand*{\Authand}{, }
\renewcommand*{\Authands}{, }
\renewcommand*{\Affilfont}{\normalsize\normalfont}
\renewcommand*{\Authfont}{\bfseries}
\setlength{\affilsep}{2em}

\author[1\thanks{shun.zhang@uvm.edu}]{Shun~Zhang}
\author[1,2]{Tabia~Tanzin~Prama}
\author[1,3]{Christopher~M.~Danforth}
\author[1,2,4,5\thanks{peter.dodds@uvm.edu}]{Peter~Sheridan~Dodds}

\affil[1]{
  Computational Story Lab,
  Vermont Complex Systems Institute,
  MassMutual Center of Excellence for Complex Systems and Data Science,
  Vermont Advanced Computing Center,
  University of Vermont,
  Burlington,
  VT 05405,
  US
}
\affil[2]{
  Department of Computer Science,
  University of Vermont,
  Burlington,
  VT 05405,
  US
}
\affil[3]{
  Department of Mathematics \& Statistics,
  University of Vermont,
  Burlington,
  VT 05405,
  US
}
\affil[4]{
  Santa Fe Institute,
  1399 Hyde Park Rd,
  Santa Fe,
  NM 87501,
  US
}
\affil[5]{
  Complexity Science Hub,
  Metternichgasse 8,
  1030 Vienna,
  Austria
}

\date{\today}

\maketitle


\mbox{}

\bigskip
\bigskip
\bigskip

\hspace*{50pt}
\begin{minipage}{300pt}

  \begin{tabular}{>{\raggedright\arraybackslash}p{180pt}p{30pt}p{180pt}}

 \parbox{180pt}{    
   \begin{loglinebox}[Logline]
     \raggedright
     We apply the Power–Danger framework of ousiometry to the ensemble sitcom `Friends' to reveal how affective meaning in dialogue evolves across individual characters, differentiates interactional contexts, and forms recurrent temporal patterns within an ensemble narrative.
     \smallskip
   \end{loglinebox}
 }

 &
    
    &
    
     \parbox{180pt}{    
    \includegraphics[width=180pt,valign=m,frame]{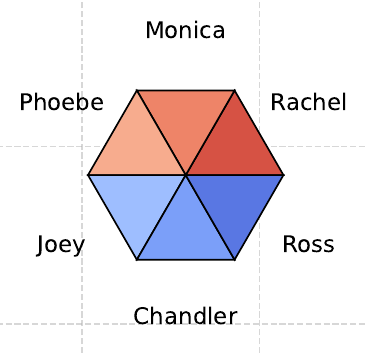}
 }
    
  \end{tabular}
  
\end{minipage}

\clearpage

\newgeometry{
  left=2in,
  right=2in,
  top=0.65in,
  bottom=0.65in,
  }

\onecolumn

\renewcommand{\baselinestretch}{1.2}
\selectfont


\begin{abstractbox}[Abstract]
  \raggedright

Computational narrative analysis increasingly seeks representations that capture how language reflects changing social and interactional contexts. We apply Ousiometry, a lexicon-based framework organized around Power, Danger, and Structure to the complete dialogue of \textit{Friends} (236 episodes, $\sim$750K tokens), examining temporal variation, group-level positioning, and pairwise co-variation among the six main characters. We find substantial episode-to-episode fluctuation in character
trajectories, with PELT-detected change points descriptively aligning with major
narrative events. At the group level, character combinations occupy different
regions of Power--Danger space, with greater dispersion in Power than Danger,
although these differences do not survive Bonferroni correction and trajectory
clustering shows weak structure (silhouette $=0.27$). Pairwise Pearson
correlations reveal modest co-variation across characters, with three
significant pairs in Power, two in Danger, and six in Structure after
correction. Partial correlations substantially reduce these associations,
leaving only one significant pair in Power and one in Danger, suggesting that
much of the observed co-variation reflects ensemble-wide rather than
pair-specific dynamics. Overall, our findings show that episode-level Ousiometry captures
meaningful temporal variation in character dialogue, while evidence for stable
group-level structure and pair-specific co-variation remains limited. Semantic
variation in \textit{Friends} is therefore better characterized as continuous
and context-dependent.

  \smallskip
\end{abstractbox}


\begin{infobox}[Keywords]
  \centering
  Ousiometry,
Computational Narrative Analysis,
Semantic Lexicon,
Power–Danger-Structure Space,
Character Interaction Dynamics,
Ensemble Sitcom,
Friends
 \smallskip
\end{infobox}

\renewcommand{\baselinestretch}{1}
\selectfont

\twocolumn

\restoregeometry

\clearpage

\tableofcontents

\clearpage

\section{Introduction}

Sentiment analysis has established itself as a standard tool for extracting affective meaning from text~\cite{Taboada2016SentimentAA}, with widespread applications in social sciences, public health, and computational narrative analysis~\cite{Taboada2016SentimentAA, Mohammad2016SentimentAD}. Most sentiment analysis systems operate by reference to a semantic lexicon: a structured inventory of words paired with affective annotations~\cite{Stone1967TheGI, vanderVeen2025TheAO}. By averaging the sentiment scores assigned to the words in a given text, the system produces an estimate of that text's overall affective orientation, as received by a reader~\cite{Esuli2006SENTIWORDNETAP}.

Underlying sentiment analysis is ousiometrics: The measurement of essential meaning.
In recent work revisiting Osgood \etal's 
longstanding 
Evaluation-Potency-Activation (EPA) framework
(and its instantiation in sentimental response
as Valence-Arousal-Dominance),
essential meaning has been shown to instead
be described a three-dimensional space with
orthogonal semantic differential axes of
Power
(\semdiff{weak}{powerful}),
Danger
(\semdiff{safe}{dangerous}),
and
Structure
(\semdiff{rigid}{loose})~\cite{Dodds2026Ousiometrics}.
In the primary plane spanned by Power and Danger,
the two intercardinal dimensions are 
Goodness (Power - Danger, \semdiff{bad}{good}) 
and Aggression (Power + Danger, \semdiff{gentle}{aggressive}).
We refer to the full circumplex model as GPADS framework.

In this study, we analyze interaction dynamics in dialogue using this Power-Danger-Structure framework.
We focus mainly on the first two dimensions of Power and Danger.

We use a case study of the television sitcom Friends~\cite{friends_tv}, which offers a rich corpus for examining the intersection of emotion and social structure~\cite{friends_tv}. Airing from 1994 to 2004, the series spans ten seasons and 236 episodes, following the lives of six close friends (Rachel, Monica, Phoebe, Joey, Chandler, and Ross) living in New York City~\cite{friends_tv}. The narrative structure of Friends is organized around recurring character groupings, which allows us to examine how semantic profile shifts as the same character interacts with different partners.

This study addresses three specific research questions::

\begin{enumerate}
    \item \textbf{Individual level:} How do the Power and Danger scores of each of the six main characters evolve across the ten seasons, and how do these trajectories align with key narrative events?

    At the individual level, we track how each speaker's Power-Danger profile evolves over time, revealing essential meaning arcs that correspond to narrative development.  

    \item \textbf{Combination level:} Do different combinations occupy distinct regions of the Power-Danger space?

    We then extend the framework to compare different character combinations, examining how different groupings of characters occupy distinct regions in the semantic space.

    \item \textbf{Pairwise level:} Do characters' semantic profiles systematically co-vary across episodes?

    We also use pairwise Pearson correlations to examine whether character trajectories exhibit nonuniform co-variation.
    
\end{enumerate}

\section{Literature Review}

\subsection{Computational Character Analysis in Narrative}

Characters are fundamental to narrative structure, linking events and sustaining coherence as stories develop over time~\cite{Eder2010CharactersIF, Margolin1986THEDA}. Accordingly, computational narrative research has developed a range of methods for representing fictional characters at scale. These approaches characterize characters through their narrative roles, personality traits, dialogue, and patterns of interaction, providing complementary views of how characters function within a story~\cite{Bamman2013LearningLP, Elson2010ExtractingSN, Brahman2021LetYC}.

\subsection{Ousiometrics}

Ousiometrics provides a quantitative framework for representing the essential meaning of words along interpretable semantic dimensions~\cite{Dodds2026Ousiometrics}. The GPADS framework characterizes words along the dimensions of Goodness, Power, Aggression, Danger, and Structure, providing a multidimensional representation of semantic variation~\cite{Dodds2026Ousiometrics}. Specifically, Power and Danger form two essential semantic axes, while Goodness and Aggression correspond to the two orthogonal bisectors of the Power--Danger plane~\cite{Dodds2026Ousiometrics}. This geometry provides two complementary coordinate systems for describing the same two-dimensional semantic space: words can be positioned in terms of Power and Danger, or equivalently interpreted along the rotated Goodness and Aggression directions. Structure provides an additional dimension of semantic organization.

An important empirical finding of Ousiometrics relates to the distribution of language within the semantic space. When words are weighted across real-world corpora, language does not occupy the Power--Danger plane symmetrically but instead exhibits a pronounced bias toward the low-Danger, or safe, region~\cite{Dodds2026Ousiometrics}. This finding provides a geometric reinterpretation of the well-known positivity bias in human language, commonly associated with the Pollyanna Principle~\cite{boucher1969a,dodds2015a}. Because the \semdiff{bad}{good} axis is a bisector within the Power--Danger plane, the observed positivity bias can be understood as a one-dimensional projection of the broader safety bias onto the Goodness dimension~\cite{Dodds2026Ousiometrics}. 

Ousiometric dimensions can be aggregated over sequential units of text to examine how semantic position changes over narrative time. Fudolig et al.~\cite{Fudolig2022ADO} applied Ousiometric measurements to books and represented semantic fluctuations as trajectories over cumulative word-time, demonstrating how changes in semantic dimensions can reveal large-scale variation in narrative structure.

\subsection{Sitcom Dialogue}

Sitcom dialogue offers distinct advantages for studying semantic profile in social interaction. It features a stable cast of recurring characters whose relationships evolve gradually over extended periods where the same individuals repeatedly interact in different social configurations~\cite{Bost2016NarrativeSD, Bamman2013LearningLP}. These long-form narratives are particularly suitable for such analysis because a recurring cast appears across many episodes, allowing character behavior to be observed repeatedly over an extended narrative sequence~\cite{Mills2009TheS}.

Among ensemble sitcoms, Friends provides an especially suitable corpus~\cite{friends_tv, Quaglio2009TelevisionDT}. The series maintains a consistent set of six central characters (Rachel, Monica, Phoebe, Joey, Chandler, and Ross)~\cite{friends_tv} across its entire ten-season run, with each episode structured around interactions drawn from the same social ensemble. The recurring combinations of Rachel, Monica, Phoebe, Joey, Chandler, and Ross create a rich range of conversational settings, from one-to-one to full-group discussions. The corpus contains approximately 750,000 tokens of dialogue, sufficient to support quantitative analysis (Corpus Details in the following Data Section). Previous corpus-based comparisons of \textit{Friends} with naturally occurring conversation have shown that its scripted dialogue shares many features with natural conversation while also exhibiting systematic linguistic differences~\cite{Quaglio2009TelevisionDT}.

\subsection{Research Gap}

Previous applications of Ousiometrics have demonstrated that semantic scores can be tracked across narrative time~\cite{Fudolig2022ADO}, allowing changes in semantic position to be represented as trajectories. Existing narrative applications have largely examined semantic variation at the level of texts or narrative segments, while the parallel semantic trajectories of recurring characters within the same narrative ensemble remain comparatively underexplored. 

This is important for ensemble narratives, where multiple recurring characters participate in a shared narrative context while maintaining distinct patterns of dialogue and interaction. Character-level Ousiometric analysis therefore offers an opportunity to extend existing trajectory-based approaches from aggregate narrative variation to semantic variation among recurring members of an ensemble. The relational analysis further distinguishes between marginal and conditional co-variation. Pairwise correlations extend existing character-trajectory comparisons to ousiometric dimensions.

\section{Data and Methods}

\subsection{Data Collection and Pre-processing}

The dataset comprises complete transcripts for all ten seasons of Friends, obtained from publicly available sources~\cite{corbari_friends_scripts}. The raw transcripts contain approximately one million word tokens across $236$ episodes; after the preprocessing steps described below, the analyzed corpus contains 747,805 tokens, an average of roughly 3,200 tokens per episode. Each transcript provides line-by-line dialogue with speaker attributions and embedded stage directions. In some seasons, the last two episodes are combined into a single file for plot coherence. They are all divided into the correct episode divisions according to IMDb~\cite{imdb_friends_episodes}.

Preprocessing proceeds through several steps. Raw text is tokenized and lowercased. Stop words are removed for the primary Ousiometric analyses described below.

Speaker aliases and abbreviated forms, such as ``RACH'' for Rachel, ``CHAN'' for Chandler, ``MNCA'' for Monica are all standardized to their full character names. Non-lexical markers, including laughter notations, are all removed. For example, in S02E10, we replace ``MNCA'' with ``Monica''.
\begin{quote}
MNCA: You were so amazing as the king. I was really impressed, I was.
\end{quote}

A key challenge in preparing the dataset for lexicon-based modeling is the lack of consistent and standardized scene annotations in the raw transcripts. The original transcripts are provided as line-by-line dialogue data but do not consistently include structured scene-level annotations. Since scene boundaries are important for analyzing interaction patterns and semantic shifts, we construct a preprocessing pipeline to segment episode scripts into discrete scenes. Scene boundaries are identified using contextual cues such as stage directions and location descriptions embedded in the transcripts.

While this automatic segmentation approach is sufficient for most episodes, we observe inconsistencies in several Season 2 episodes, where scene transitions are not all explicitly marked in the source data. To ensure the reliability of scene-based analysis, we manually annotate scene boundaries for ten episodes of Season 2. This manual annotation process involves carefully examining the narrative flow of each episode and inserting standardized scene markers at appropriate transition points. The corrected annotations are used consistently in all subsequent analyses. For example, in S02E22 ``The One with the Two Parties'',
\begin{quote}
[Back at Chandler and Joey's party. Everyone is dancing and having fun.]
\end{quote}
is corrected into
\begin{quote}
[Scene: Back at Chandler and Joey's party. Everyone is dancing and having fun.]
\end{quote}

After preprocessing, we have the distribution of all six main characters' tokens across all episodes in Table~\ref{tab:corpus_statistics}.

\begin{table}[ht!]
\centering
\begin{tabular}{lcc}
\toprule
\hline
\textbf{Corpus subset} & \textbf{Unique tokens} & \textbf{Total tokens} \\
\hline
\midrule
Rachel   & 6,390 & 106,735 \\
Monica   & 6,022 & 90,411  \\
Phoebe   & 6,492 & 90,400  \\
Joey     & 6,645 & 99,609 \\
Chandler & 6,896 & 96,234 \\
Ross     & 7,182 & 107,063 \\
Others   & 9,291 & 157,353 \\
\midrule
\textbf{Entire series} & \textbf{20,049} & \textbf{747,805} \\
\hline
\bottomrule
\end{tabular}
\caption{Descriptive statistics of the Friends dialogue corpus.
Unique tokens denote the number of distinct word forms after tokenization,
while total tokens denote the total number of word occurrences.}
\label{tab:corpus_statistics}
\end{table}

The cleaned corpus contains 747,805 word tokens and 20,049 unique tokens across all ten seasons of \textit{Friends}. Our character-level analyses focus on the six main characters: Rachel, Monica, Phoebe, Joey, Chandler, and Ross. Together, their dialogue accounts for approximately 78.96\% of the full corpus.

\subsection{Ousiometrics Framework}

In RQ1 and RQ2, we mainly focus on Power and Danger because they are the two primary dimensions of the ousiometric space and provide the most interpretable representation of social interaction patterns among character combinations. Each word in the lexicon thus carries two continuous values: a Power score and a Danger score. Descriptive statistics for the Power–Danger–Structure lexicon are shown in Table~\ref{tab:ousio_stats}.

\begin{table}[ht!]
\centering
\begin{tabular}{lccccc}
\toprule
\textbf{Dimension} & \textbf{Lexicon} & \textbf{Mean} & \textbf{SD} & \textbf{Min} & \textbf{Max} \\
\midrule
Power  & 20,006 & 0.000 & 0.218 & -0.612 & 0.758 \\
Danger & 20,006 & 0.000 & 0.218 & -0.591 & 0.681 \\
Structure & 20,006 & 0.000 & 0.097 & -0.382 & 0.432 \\
\bottomrule
\end{tabular}
\caption{Descriptive statistics of the Power--Danger--Structure lexicon used in this study. }
\label{tab:ousio_stats}
\end{table}

Based on this lexicon, the coverage rate for the Friends corpus is thus shown in Table~\ref{tab:ousio_coverage}. 

\begin{table}[ht!]
\centering
\begin{tabular}{lccc}
\hline
& Unique Tokens 
& Unique in Lexicon 
& Unique Coverage \\
\hline
& 20,049 
& 7,756 
& 38.69\% \\
\hline
& Total Tokens 
& Total in Lexicon 
& Total Coverage \\
\hline
& 747,805
& 222,786 
& 29.79\% \\
\hline
\end{tabular}
\caption{Lexicon coverage rates for the Friends corpus based on the Power--Danger--Structure lexicon.}
\label{tab:ousio_coverage}
\end{table}

\subsection{RQ1: Semantic Trajectories of Main Characters}
To answer our first research question: "How do individual characters' semantic profiles evolve over time in the Power-Danger space?", we aggregate all the script data into a time series for each character. All tokens of each character in each episode are aggregated into a mean power score and a mean danger score.
\begin{equation}
P_{c,t}=\frac{1}{N_{c,t}}\sum_{i=1}^{N_{c,t}} p_i,
\end{equation}

\begin{equation}
D_{c,t}=\frac{1}{N_{c,t}}\sum_{i=1}^{N_{c,t}} d_i,
\end{equation}

where $N_{c,t}$ is the number of dialogue tokens spoken by character $c$ in episode $t$, and $p_i$ and $d_i$ denote the Power and Danger scores assigned to token $i$, respectively.

To better show the arcs, we employ LOESS (Locally Estimated Scatterplot Smoothing)~\cite{Cleveland1979RobustLW} smoothing with a span parameter of $10$ episodes. The smoothing is applied independently to each character's Power and Danger sequences. Let

\begin{equation}
\hat{P}_{c,t}=\mathrm{LOESS}(P_{c,t};\, s),
\end{equation}

\begin{equation}
\hat{D}_{c,t}=\mathrm{LOESS}(D_{c,t};\, s),
\end{equation}
where $s=10$ denotes the smoothing span, corresponding to a local neighborhood of ten episodes. The resulting smoothed trajectories provide an intuitive representation of each character's semantic trajectories throughout the series.

The smoothing arcs show gradual patterns. They cannot indicate statistically where a character's sentiment undergoes a structural shift.
We thus try to detect the change points. Narrative developments in serialized television are typically event-driven~\cite{Porter2002ReF}. Specific plot events, such as the beginning or end of a relationship, a career milestone, or a major life decision, can produce discrete shifts in how that character speaks and interacts. Change point detection locates the episodes where these shifts occur, allowing us to examine whether detected change points descriptively coincide with identifiable narrative developments.

To locate those points in a time series where the statistical properties of the sequence change, we apply the PELT (Pruned Exact Linear Time) algorithm~\cite{Killick2011OptimalDO} to each character's episode-level Power and Danger time series. PELT supplies such a criterion through its penalized likelihood framework: a break is retained only if the improvement in fit outweighs the cost of adding an additional segment~\cite{Killick2011OptimalDO}. This prevents the over-interpretation of minor fluctuations as narratively significant~\cite{Truong2020SelectiveRO}. Penalty is decided by sensitivity criterion. We conducted a sensitivity analysis across penalty values of $0.5$, $1.0$, $1.5$, and $2.0$. The resulting number and locations of detected change points are compared across these settings. The penalty(0.5) is selected by examining the stability of the number and locations of change points across neighboring penalty values.

\paragraph{Word-shift analysis}
As a supplementary lexical analysis, we use weighted-average word shifts to examine which words distinguish each character from the remaining five characters along the Power and Danger dimensions. For each character, the reference distribution consists of that character's dialogue, while the comparison distribution contains the dialogue of the other five main characters. Weighted-average word shifts decompose the difference between two lexical distributions into word-level contributions based on both differences in relative word frequency and each word's Power or Danger score. The words with the largest absolute contributions therefore identify the lexical items that most strongly differentiate the two distributions along the corresponding semantic dimension. The unit of word-shift analysis here is different from the episode-level units of primary ousiometry analysis.

\subsection{RQ2: Sentiment Patterns in Character Combinations}
\begin{figure*}[t!]
    \centering
    \includegraphics[width=\textwidth]{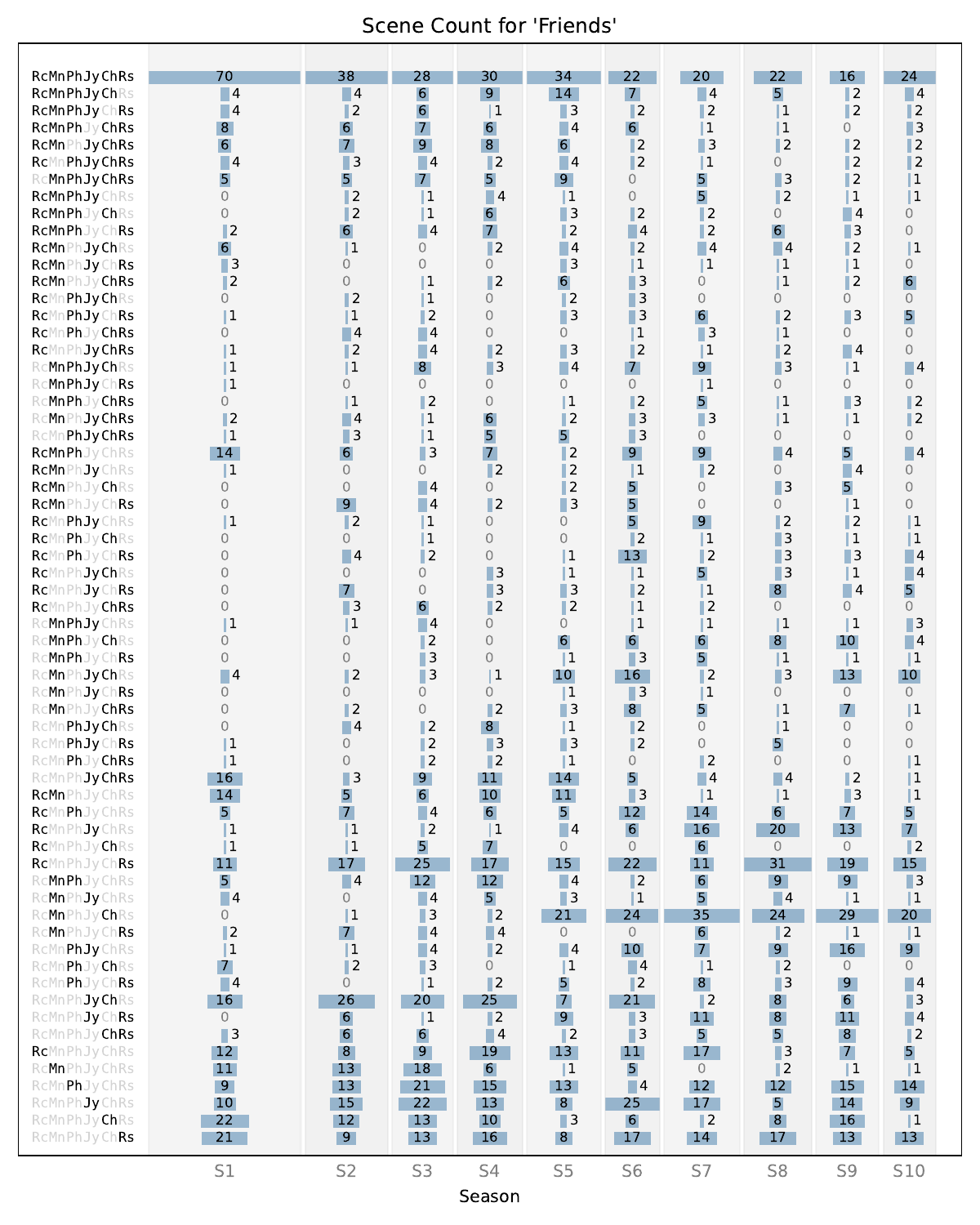}
    \caption{Character Combination Participation Across Seasons in Friends. The horizontal bar chart displays the number of scenes per season for all 63 non-empty character combinations. Seasons S1--S10) are aligned along the x-axis, with bar lengths scaled to the number of scenes featuring each combination per season.
    }
    \label{fig:scene_count}
\end{figure*}

\begin{figure*}[t!]
    \centering
    \includegraphics[width=\textwidth]{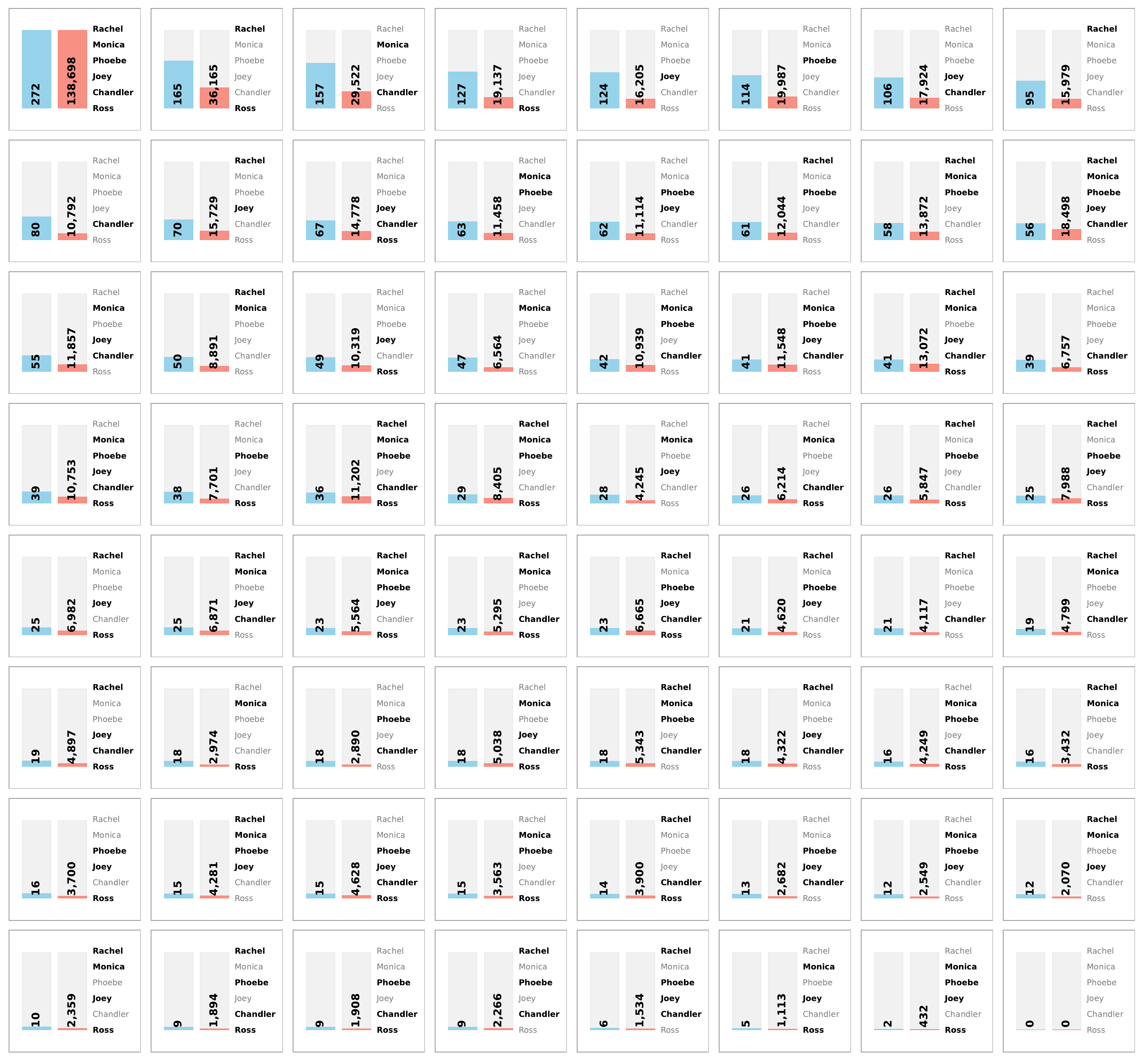}

    \caption{
Scene count and word count per character combination. Blue bars represent
scene counts, while red bars represent word counts. Going across and then
down, panels are arranged in descending order of scene count.
}
    \label{fig:side_bar_plot_scene_word_count}
\end{figure*}

We conduct a combination-level analysis of the Power-Danger semantic space. With six main characters, there exist 63 non-empty subsets, each representing a distinct narrative structure. Figure~\ref{fig:scene_count} shows the total scene counts for all 63 combinations in the show through all ten seasons.
Figure~\ref{fig:side_bar_plot_scene_word_count} shows both scene count and word count for all combinations in descending order of scene counts.

For each combination, we aggregate Power and Danger scores across all scenes in which exactly that set of characters appears together.

  \begin{figure*}[ht!]
      \centering
      \includegraphics[width=0.75\linewidth]{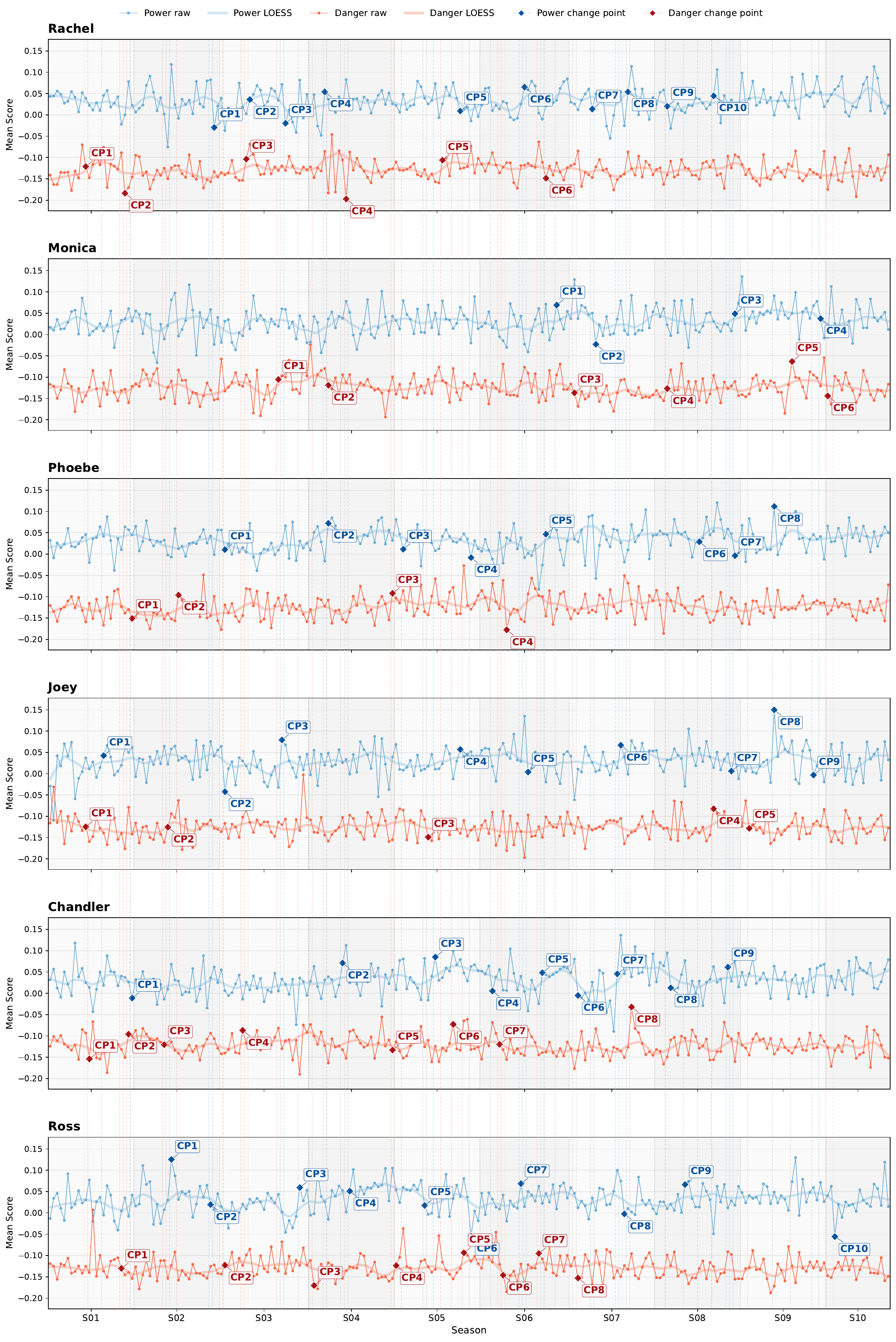}
      \caption{Trajectories of Power (blue) and Danger (red) for the six main Friends characters across all episodes. Raw episode-level mean scores are shown as solid dots and lines, while LOESS-smoothed curves summarize long-term trends. Detected change points indicate major shifts in the corresponding Power or Danger trajectories identified by the change-point detection algorithm for each character. Vertical gray bands denote season boundaries. Details for all Change Points are in Appendix~\ref{Appendix:change_point}.}
      \label{fig:friends_all_characters_power_danger_time_series_loess}
    \end{figure*}

\paragraph{Data Filtering}
Character combinations vary dramatically in frequency throughout the series. Some dyads (two-character groups) and the full ensemble appear hundreds of times, while certain groups only appear in fewer than 10 scenes. We impose two filtering criteria to ensure the statistical reliability of the estimates of the power and danger scores. A combination is retained for analysis if it meets both of the following conditions: (1) it appeared in at least 10 scenes, and (2) the total number of words spoken by the characters across all those scenes is at least 2,000. The criterion ensures sufficient observations to obtain stable and meaningful results from the semantic space. These thresholds are informed by the distribution of scene counts and word counts across all 63 combinations, as shown in Figure~\ref {fig:scene_count} and Figure~\ref{fig:side_bar_plot_scene_word_count}. The filtering procedure yielded 56 combinations for subsequent analysis, representing a balance between comprehensive coverage and statistical reliability. In Appendix~\ref{Appendix:combination_remain}, Table~\ref{table:combination_filter} shows the seven combinations that have been removed for further analysis, and Table~\ref{table:combination_remain} shows that all 56 combinations remain for the analysis.

\paragraph{Combination Score Calculation:}
For each combination $g$, let $\mathcal{S}_g$ denote the set of scenes in which $g$ appears. For each scene $s \in \mathcal{S}_g$, let $P_s$ and $D_s$ be the mean Power and Danger scores of all utterances in that scene. The aggregate score for combination $g$ is the average of these scene-level values:
\[
\bar{P}_g = \frac{1}{|\mathcal{S}_g|}\sum_{s \in \mathcal{S}_g} P_s, \qquad
\bar{D}_g = \frac{1}{|\mathcal{S}_g|}\sum_{s \in \mathcal{S}_g} D_s
\]

\paragraph{Bootstrap Confidence Ellipses:}

To assess the statistical significance of differences in position among combinations in the Power-Danger space, we employed a nonparametric bootstrap procedure. For each combination $g$ with $n_g = |\mathcal{S}_g|$ observed scenes, we generated $B = 10{,}000$ bootstrap samples by resampling with replacement from its set of scene-level Power-Danger coordinates $\{(P_s, D_s)\}_{s \in \mathcal{S}_g}$. Each bootstrap sample produced a mean vector $\bar{\mathbf{X}}_g^{*(b)} = (\bar{P}_g^{*(b)}, \bar{D}_g^{*(b)})$, giving a bootstrap distribution of the combination's estimated semantic position.

The 95\% confidence ellipse for combination $g$ is then constructed from the covariance matrix of the bootstrap means:

\[
\hat{\Sigma}_g = \frac{1}{B-1}\sum_{b=1}^{B}\left(\bar{\mathbf{X}}_g^{*(b)} - \bar{\mathbf{X}}_g\right)\left(\bar{\mathbf{X}}_g^{*(b)} - \bar{\mathbf{X}}_g\right)^\top
\]

where $\bar{\mathbf{X}}_g = (\bar{P}_g, \bar{D}_g)$ is the observed mean. The confidence region is defined by all points $\mathbf{x} = (P, D)$ satisfying:

\[
(\mathbf{x} - \bar{\mathbf{X}}_g)^\top \hat{\Sigma}_g^{-1} (\mathbf{x} - \bar{\mathbf{X}}_g) \leq \chi^2_{2, 0.95}
\]

where $\chi^2_{2, 0.95} \approx 5.99$ is the 95th percentile of the chi-square distribution with two degrees of freedom. This ellipse captures the region within which the true population mean of the combination's Power-Danger position is expected to fall with 95\% confidence.

The area of each confidence ellipse, computed as $A_g = \pi \cdot \chi^2_{2,0.95} \cdot \sqrt{\det(\hat{\Sigma}_g)}$, serves as an index of estimation precision. That is to say, combinations appearing in more scenes produce smaller ellipses and thus more precisely estimated positions.

We also need to define a global centroid to determine which combinations occupy positions that are statistically distinct from the whole group. Let $\bar{\mathbf{X}}_{\text{global}} = (\bar{P}_{\text{global}}, \bar{D}_{\text{global}})$ denote the mean of all 56 retained combination means. This centroid represents the average semantic position within the ensemble in the power-danger space. Thus, a combination $g$ is classified as significantly deviating from the global centroid if its 95\% confidence ellipse does not contain $\bar{\mathbf{X}}_{\text{global}}$. To characterize the direction of deviation, we classify those significant combinations into four quadrants relative to the global centroid: high Power $+$ low Danger ($P > P_{\text{global}}$, $D < D_{\text{global}}$), high Power $+$ high Danger ($P > P_{\text{global}}$, $D > D_{\text{global}}$), low Power $+$ low Danger ($P < P_{\text{global}}$, $D < D_{\text{global}}$), and low Power $+$ high Danger ($P < P_{\text{global}}$, $D > D_{\text{global}}$).

The combination-level semantic space is visualized using a two-dimensional scatter plot. Each retained combination appears as a point at its observed mean coordinates $(\bar{P}_g, \bar{D}_g)$. The 95\% bootstrap confidence ellipse is overlaid around each point, with ellipse area scaled to indicate estimation precision. The global centroid is marked as a reference point. Combinations whose ellipses do not contain the global centroid are highlighted and labeled.

\paragraph{Trajectory Clustering Analysis of Character Combinations:}

To capture the dynamic patterns in combinational trajectories, we examine how each combination's Power-Danger position shifts across the ten-season narrative timeline. Since RQ1 has already established that individual characters change through the series, we now determine whether combinations exhibit any patterns of change that can be grouped into interpretable trajectory types.

We apply trajectory clustering to the seasonal Power-Danger sequences of the character combinations. This analysis proceeds from the 56 character combinations identified. For each combination, we aggregate Power and Danger scores by season, getting a set of seasonal mean values $\bar{P}_{g,s}$ and $\bar{D}_{g,s}$ for each combination $g$ in season $s (s=1,2,...,10)$. To maintain trajectory stability, we then impose two inclusion criteria. A season is considered valid for a given combination if the combination appears in at least three scenes during that season. A combination is retained for trajectory analysis if it satisfies the first criterion in at least eight of the ten seasons. Of the 56 combinations examined in RQ2, 15 meet these two criteria as shown in Table~\ref{table:cluster_combination} in Appendix~\ref{Appendix:Character_Combinations}.

For the retained combinations, missing seasonal values are imputed using linear interpolation to ensure that all trajectories are represented as complete 10-season sequences. Endpoint values, where interpolation was not possible, are carried forward or backward from the nearest observed season. Each trajectory is thus encoded as a 10-dimensional sequence $T_g = \{ \bar{P}_{g,s}, \bar{D}_{g,s} \}^{10}_{s=1}$, where each time point is a two-dimensional vector of Power and Danger scores. Prior to distance calculation, Power and Danger scores are standardized using z-score normalization.

Trajectory similarity is assessed using multivariate Dynamic Time Warping (DTW)~\cite{Sakoe1978DynamicPA, Berndt1994UsingDT}. A Sakoe-Chiba band of width two seasons is applied to constrain the alignment~\cite{Sakoe1978DynamicPA}.

Hierarchical clustering is then performed using the complete linkage method on the DTW distance matrix~\cite{Murtagh2012AlgorithmsFH}. The number of clusters is selected by jointly considering the silhouette coefficient across candidate solutions from 2 to 10 clusters~\cite{Rousseeuw1987SilhouettesAG}.

\subsection{RQ3: Do characters' semantic profiles systematically co-vary across episodes?}

To examine whether characters exhibit coordinated semantic variation over time, we use the original episode-level Power, Danger, and Structure (PDS) time series. This analysis focuses on whether main characters' episode-to-episode fluctuations co-vary. For each dimension, Pearson correlations are calculated between the time series of every pair of the six main characters. With six characters, this produces $\binom{6}{2}=15$ pairwise correlations per dimension. A positive correlation indicates that two characters tend to move in the same direction relative to their respective series-level means across episodes, whereas a negative correlation indicates opposing co-variation. Statistical significance is assessed using two-sided tests of $H_0:\rho=0$, with Bonferroni correction applied to the 15 pairwise comparisons within each dimension.

Because episode-level ousiometric scores are estimated from finite samples of dialogue tokens, we additionally assess the reliability of the PDS time series using split-half reliability~\cite{Spearman1910CORRELATIONCF, Brown1910SER}. For each character in each episode, the lexicon-matched tokens are randomly divided into two approximately equal subsets. Power, Danger, and Structure scores are calculated independently for the two halves, producing two episode-level time series for each character and dimension. The two half-series are then correlated across episodes using Pearson's $r$. Because each correlation is based on half of the available tokens, the resulting coefficient is adjusted using the Spearman--Brown correction~\cite{Spearman1910CORRELATIONCF, Brown1910SER},

\begin{equation}
r_{\mathrm{SB}}
=
\frac{2r_{\mathrm{half}}}
{1+r_{\mathrm{half}}},
\end{equation}

where $r_{\mathrm{half}}$ is the correlation between the two independently scored halves. This produces 18 reliability estimates corresponding to six characters across the three PDS dimensions and provides an estimate of the measurement stability underlying the character-level correlation analysis.

Finally, to distinguish pair-specific co-variation from variation shared across the ensemble, we compute partial correlations between each pair of characters while controlling for the remaining four characters. Partial correlations are calculated from the covariance matrix of the original episode-level character trajectories. For each PDS dimension, let $\Sigma$ denote the $6\times6$ covariance matrix of the six character trajectories and let $\Omega=\Sigma^{-1}$ denote its precision matrix. The partial correlation between characters $i$ and $j$, conditional on the other four characters, is

\begin{equation}
r_{ij\cdot\mathrm{rest}}
=
-\frac{\Omega_{ij}}
{\sqrt{\Omega_{ii}\Omega_{jj}}}.
\end{equation}

These partial correlations quantify the residual linear association between two character trajectories after accounting for their linear relationships with the rest of the ensemble. Significance is evaluated using two-sided tests, with Bonferroni correction applied to the 15 character-pair comparisons within each dimension.

\section{Results and Discussion}

\subsection{RQ1: Sentiment Arcs of Main Characters}

\subsubsection{Overall Power-Danger Profiles}

We examine whether the six principal characters exhibit distinguishable Power-Danger profiles when aggregated across the entire series. For this analysis, we treat each episode as a blocking factor in a repeated-measures design, with character as the within-subject variable. A total of 236 episodes are included in the analysis, yielding 1,416 character-episode observations. Table~\ref{table:PD_descriptive} presents the episode-level mean Power and Danger scores for each character.

\begin{table}[h!]
\centering
\begin{tabular}{lcc}
\hline
Character & Power ($M\pm SD$) & Danger ($M\pm SD$) \\
\hline
Rachel   & $0.0325 \pm 0.0314$ & $-0.1289 \pm 0.0246$ \\
Monica   & $0.0311 \pm 0.0337$ & $-0.1234 \pm 0.0271$ \\
Phoebe   & $0.0330 \pm 0.0308$ & $-0.1222 \pm 0.0271$ \\
Joey     & $0.0288 \pm 0.0336$ & $-0.1250 \pm 0.0264$ \\
Chandler & $0.0316 \pm 0.0338$ & $-0.1215 \pm 0.0248$ \\
Ross     & $0.0323 \pm 0.0326$ & $-0.1277 \pm 0.0262$ \\
\hline
\end{tabular}
\caption{Descriptive statistics of Power and Danger scores for the six main characters.}
\label{table:PD_descriptive}
\end{table}

To test whether the observed descriptive differences reach statistical significance, we conduct repeated-measures ANOVAs for Power and Danger, treating episode as the repeated-measures unit and character as the within-subject factor. Mauchly's test indicates no evidence of violations of the sphericity assumption for either Power or Danger. 

For Power, the main effect of character is not significant with $F(5, 1175) = 0.587, p = 0.710$, indicating no evidence of systematic differences in episode-level Power scores across the six main characters. In contrast, the effect of character on Danger is statistically significant with $F(5,1175)=3.568, p=.003$, although the associated effect size is small $(\eta_p^2=.015)$. 

We conduct paired \(t\)-tests for all 15 character pairs separately for Power and Danger, applying a Bonferroni correction within each outcome. No pairwise differences in Power were significant after correction. For Danger, only the Rachel--Chandler comparison remained significant after Bonferroni correction, $t(235)=-3.421, p_{\mathrm{adj}}=.011$. Chandler had a slightly higher mean Danger score $(M=-.1215)$ than Rachel $(M=-.1289)$. No other pairwise differences in Danger survive the correction.

Overall, the six characters occupy a broadly similar region of Power--Danger space at the series level.

Notably, all six characters have positive mean Power scores and negative mean Danger scores, placing them in a common powerful-safe region of the space. In ousiometric terms, this shared baseline may reflect the common conversational environment in which the characters repeatedly participate across the series. As an ensemble sitcom, Friends places its characters in recurring social settings and conversational interactions, many of which involve forms of conversational humor such as irony and other humor-oriented exchanges~\cite{Dynel2009BeyondAJ}. These shared narrative and conversational contexts may contribute to the broadly similar powerful-safe region occupied by the six main characters.

\subsubsection{Lexical Differences: Word Shift Analysis}
Although the aggregate Power-Danger profiles are broadly similar, the characters still differ in the lexical patterns according to their dialogue. To examine these differences descriptively, we compare each character with the pooled dialogue of the remaining five characters using weighted-average word shifts. 

\begin{figure*}[h!]
    \centering
    \includegraphics[width=1.0\textwidth]{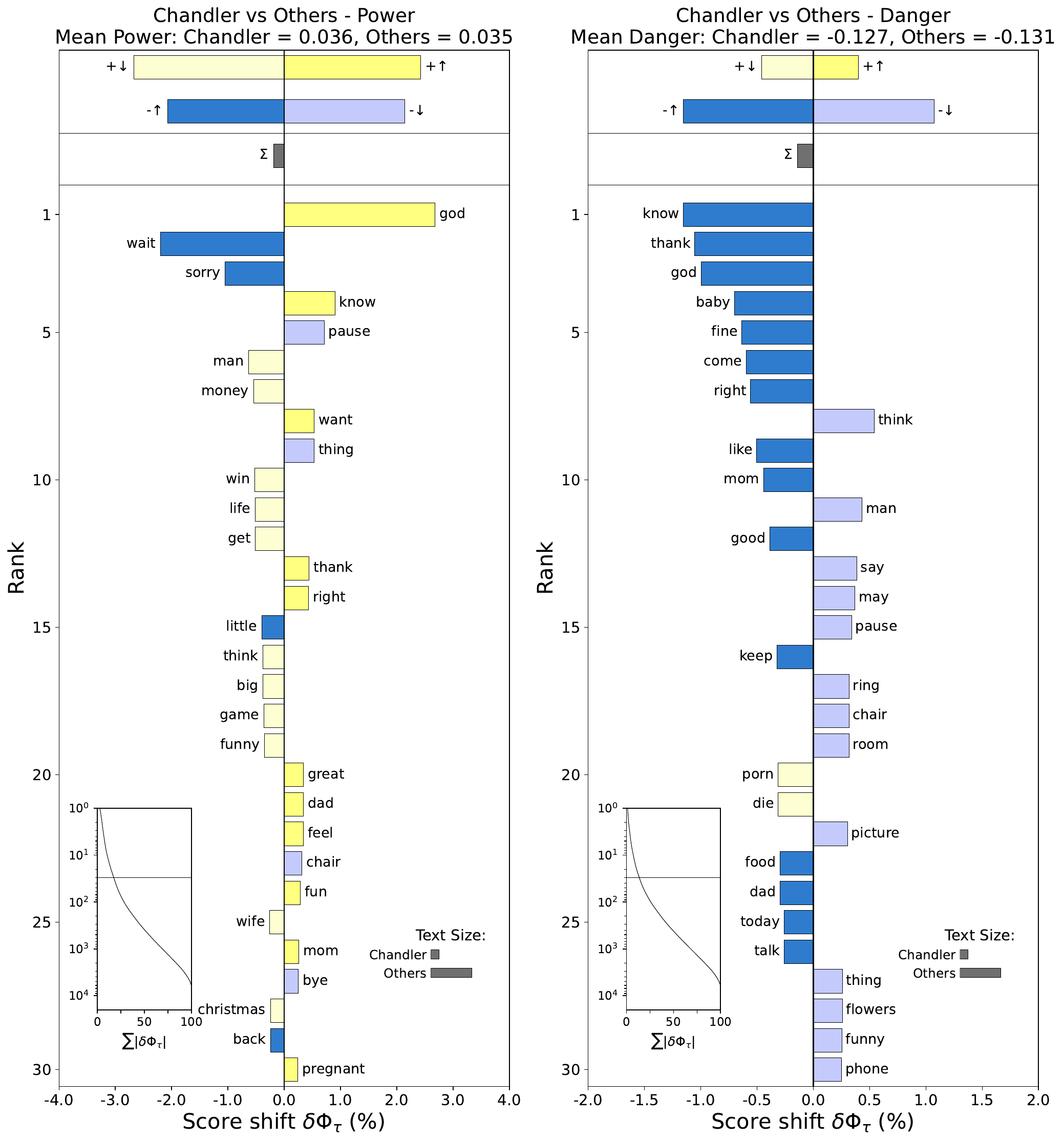}
    \caption{ Weighted-average word shift comparing the Power and Danger dimensions of Chandler's dialogue with the pooled dialogue of the other five main characters. Yellow indicates words with relatively higher power/danger scores. Blue indicates words with relatively lower power/danger scores. Darker and lighter shades indicate words used relatively more and less frequently by Chandler. Contributions to the right increase Chandler's relative power/danger, whereas contributions to the left decrease it. The legend includes the text size of both sides. }
    \label{fig:chandler_power_shift}
\end{figure*}

Figure~\ref{fig:chandler_power_shift} provides an example for Chandler's Power dimension. The word shift identifies the lexical items that contribute most strongly to the difference between Chandler's dialogue and that of the other five characters. For Chandler, for example, the relatively frequent use of the higher-power word `god' pushes his power upward, whereas the relatively frequent use of lower-Power words such as wait and sorry pushes it downward. Additionally, the mean values shown in the figure are pooled, token-frequency-weighted lexical means used for the word-shift decomposition and therefore differ from the episode-level means reported in the preceding analyses, where each episode constitutes the unit of analysis. The shift therefore reveals lexical contributions that are obscured by the aggregate Power score.

\subsubsection{Sentiment Arcs for Main Characters}
To capture the narrative dynamics underlying these static profiles, we examine how each character's Power and Danger scores evolved over the 236-episode timeline. For each episode, we compute the mean Power and Danger scores per character based on all dialogues in that episode. These raw episode-level sequences capture the full temporal resolution of the data but exhibit substantial short-term variability due to episode-specific topics, comedic subplots, and incidental interactions. To extract the underlying narrative signal, we apply LOESS (Locally Estimated Scatterplot Smoothing)~\cite{Cleveland1979RobustLW} to each character's Power and Danger sequences. The smoothing parameter (span = 10 episodes) is selected to preserve seasonal trends. Figure~\ref{fig:friends_all_characters_power_danger_time_series_loess} presents the trajectories and the smoothed curve for all six characters.

\paragraph{Trends Interpretation:}

We observe that each character possesses some shape characteristics. For example, Rachel's Danger trajectory is relatively elevated during parts of the first season and subsequently declines. This temporal pattern broadly coincides with her early narrative transition toward greater independence~\cite{wikipedia_Rachel_Green, friends_tv}. For Monica, the power score trends upward from Season 5 through Season 8, with a second increase in Season 9, which is a period that overlaps with the development of her relationship and marriage with Chandler~\cite{wikipedia_Monica_Geller, friends_tv}. For Phoebe, there is a low point in power score in Season 4, when it reaches a local minimum. Joey demonstrates the least variation across both dimensions, with scores remaining near baseline throughout the series. Chandler presents the most pronounced positive Power trend, accelerating between Seasons 5 and 7. From those observations, these trajectories differ descriptively in the direction and magnitude of their temporal variation.

While LOESS smoothing reveals continuous trends, it does not identify discrete moments at which affective trajectories undergo structural shifts. To detect such change points where episodes mark significant changes in Power or Danger score, we applied the PELT (Pruned Exact Linear Time) algorithm to each character's time series. Table~\ref{Tab:change_point_power} and Table~\ref{Tab:change_point_danger} in Appendix~\ref{Appendix:change_point} lists all detected change points with their corresponding episode.

\paragraph{Possible Change Point Interpretation}

The detected change points(CP) also show descriptive correspondence with major narrative events in the show. Chandler's power trajectory provides an illustrative example as shown in Figure~\ref{fig:friends_all_characters_power_danger_time_series_loess}. Most of his detected change points in power occur near transitions in his interpersonal and romantic storylines. Several of these change points occur during the middle seasons of the series, particularly between Seasons 4 and 8, when Chandler's major relationship storyline with Monica develops. For example, CP2 (S04E10) occurs during his relationship with Kathy, while CP3 (S05E12) falls during the early stage of his relationship with Monica. Subsequent change points at S06E04, S06E18, S07E03, and S07E14 occur during the progression of the Chandler--Monica relationship toward greater commitment, including the period surrounding their engagement. This proximity to major narrative developments therefore suggests that some changes in Chandler's lexical Power profile coincide with transitions in his long-term relationship arc. Additional descriptive support for these interpretations can be found in the change-point patterns of other characters. In particular, several changes identified in Chandler's trajectory occur near change points independently detected for Monica during periods associated with their shared relationship arc (CP1,2,3 for Monica in power).

\subsection{RQ2: Sentiment Patterns in Character Combinations}

\subsubsection{Spatial Distribution of Character Combinations}

We next examine how different character groupings occupy distinct regions in the Power-Danger semantic space. For each of 56 qualifying combinations, we compute the mean Power and Danger scores across all scenes in which that exact set of characters appeared. This produced a spatial map of the ensemble's structure, in which each combination occupies a coordinate that reflects its characteristic interpersonal stance.

\begin{figure*}[ht!]
    \centering
    \includegraphics[width=\textwidth]{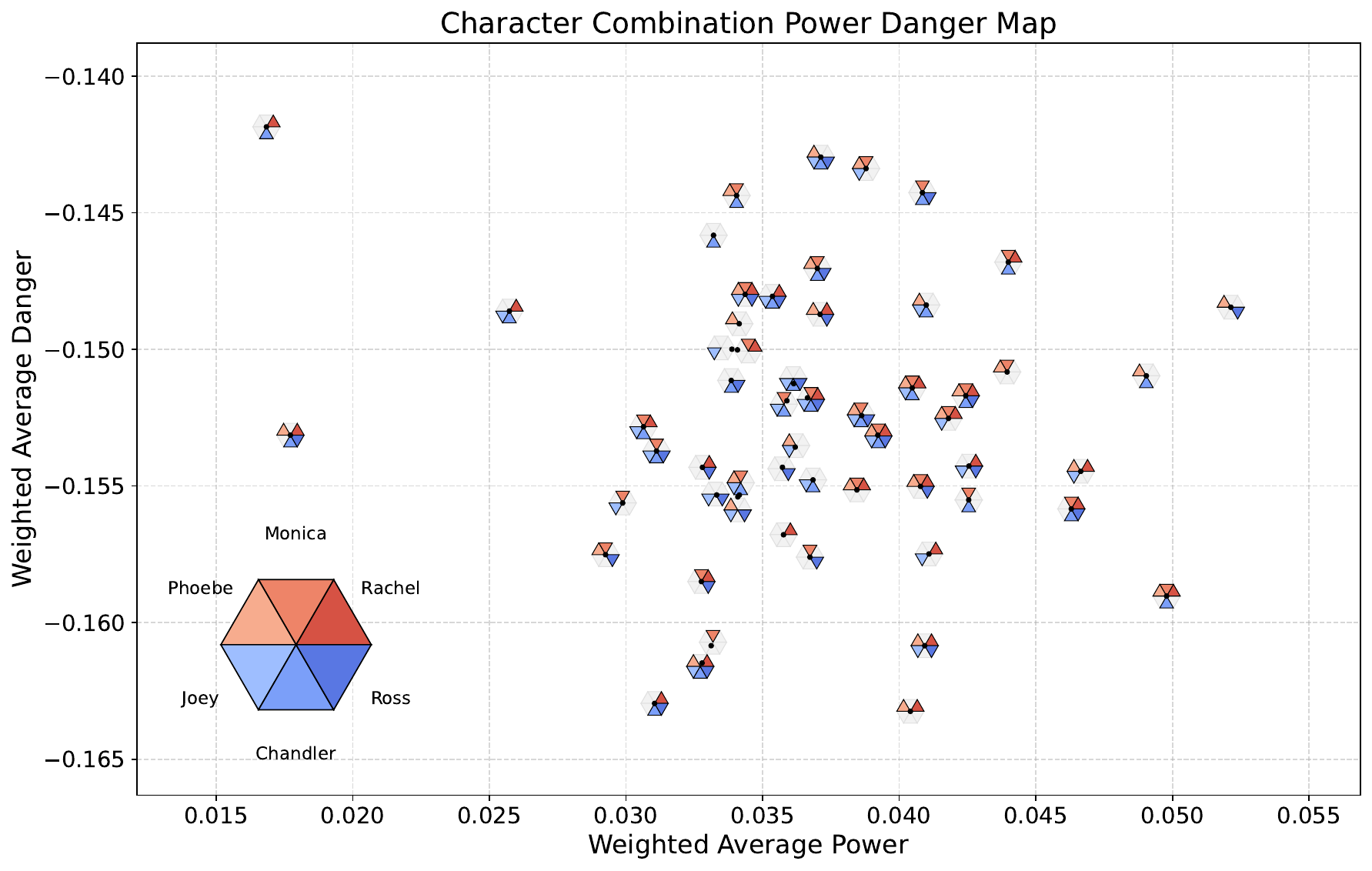}
    \caption{Mean Power--Danger positions of the 56 retained character combinations. Each hexagon represents one character combination, using different colors to distinguish six different characters. The coordinates of each center point of the hexagon mark the mean power score and the mean danger score. Small positional offsets are applied to overlapping markers for visualization only. }
    \label{fig:hexagon_plot}
\end{figure*}

Figure~\ref{fig:hexagon_plot} visualizes the power-danger position of all retained character combinations. In Figure~\ref{fig:hexagon_plot}, each combination is shown as a hexagon in the power-danger space. The center of each hexagon represents the average value for that combination, while the hexagon's color indicates what combination it represents. It provides a clear and intuitive overview of how different character groupings occupy different emotional regions in the Power-Danger space.

The spatial distribution reveals that, first, combinations are concentrated in the lower half of the Danger dimension---only two combinations exhibit Danger scores above $-0.14$, and none approach zero or a positive Danger score. This systematic bias toward low Danger across all groups suggests that the show's dialogue, regardless of which characters are interacting, maintains a baseline of interpersonal safety consistent with its everyday comedic ensemble tone. This baseline should not be read as a property of the show without further evidence, however, because the lexicon's zero is the unweighted mean over word types and ordinary frequency-weighted English already falls on the safe side of it. But whether this pattern reflects the sitcom genre more generally requires comparison with other narrative corpora.

Second, the Power dimension exhibits greater dispersion than Danger. Power scores range from 0.0137 (Rachel+Phoebe+Chandler+Ross) to 0.0566 (Phoebe+Ross), a spread of roughly four times the Danger range. Descriptively, Power varies more widely across character combinations than Danger, suggesting that Power may be more responsive to group composition in this corpus.

We also examine combinations with extreme positions on each dimension, regardless of whether their confidence ellipses excluded the centroid. These extremes identify the semantic boundaries of the ensemble in Friends. On the Power dimension, after examining the top 5 power combinations, we notice a pattern that Phoebe is in each combination. This pattern suggests that combinations involving Phoebe tend to occupy the higher end of the observed Power range. On the Danger dimension, the lowest-Danger combinations all include Rachel, and all exclude Monica. These are patterns descriptively consistent with their supportive friendship dynamic.

\subsubsection{Bootstrapped Confidence Ellipses}

Now, we apply bootstrapped 95\% confidence ellipses for each combination. For each character combination, we resample its scenes with replacement 10,000 times. Then we recalculate the mean Power and Danger coordinates for each bootstrap sample and define the confidence region using the covariance matrix of the bootstrap distribution. The result shows that five out of all 56 combinations have significant results, which means their ellipses exclude the global centroid. 

However, those results are not significant after Bonferroni correction for multiple comparisons. None of the combination survived Bonferroni correction.


\subsubsection{Trajectory Clustering of Character Combinations}

Of the 56 character combinations analyzed in RQ2, 15 met the trajectory inclusion criteria: presence in at least eight seasons with a minimum of three scenes per valid season. These 15 combinations include the most frequent groupings in the series, collectively accounting for the vast majority of ensemble dialogue scenes (Details in Table~\ref{table:cluster_combination} in Appendix~\ref{Appendix:Character_Combinations}).

Hierarchical clustering with 'complete' linkage is applied to the multivariate DTW distance matrix computed from the seasonal Power-Danger trajectories of these 15 combinations~\cite{Berndt1994UsingDT}. The silhouette coefficient is evaluated for cluster counts ranging from 2 to 10 and reaches its maximum value of 0.2668 at four clusters~\cite{Rousseeuw1987SilhouettesAG}. The maximum silhouette coefficient is only 0.2668, indicating weak cluster separation and providing little evidence for a well-defined clustering structure.

\subsection{RQ3: Do characters' semantic profiles systematically co-vary across episodes? }

\subsubsection{Character-Level Pearson Correlations}

\begin{figure*}[t!]
    \centering
    \includegraphics[width=\textwidth]{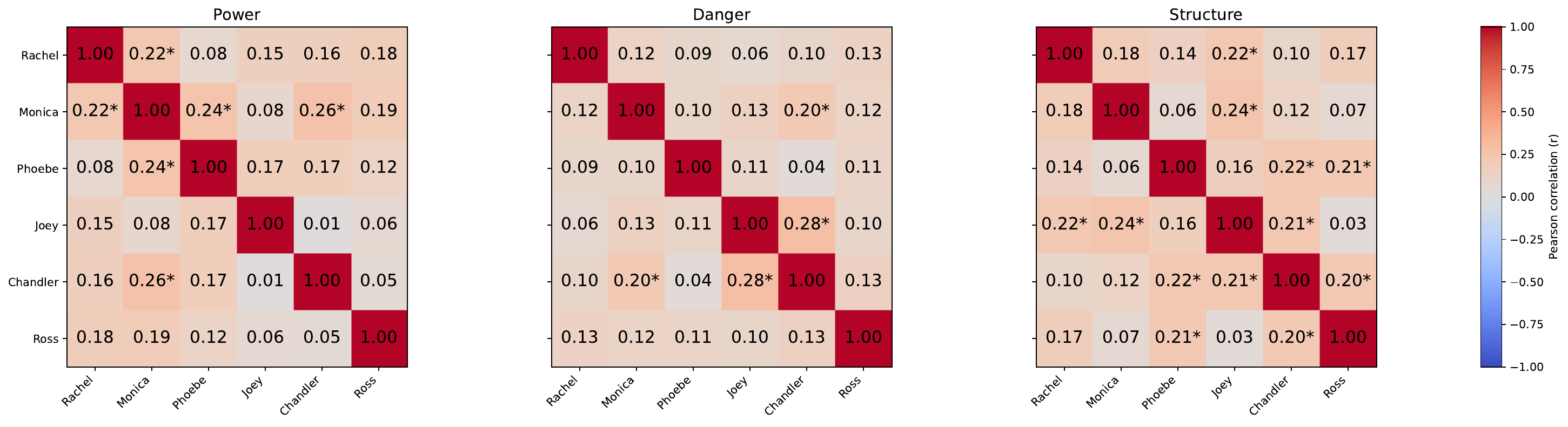}
    \caption{
    Pairwise Pearson correlations between the episode-level Power, Danger, and Structure trajectories of the six main characters which are computed from the original episode-level time series. Red indicates positive correlations, and blue indicates negative correlations. Positive values indicate that two characters tend to vary in the same direction across episodes. Asterisks indicate correlations that remain statistically significant after Bonferroni correction for the 15 pairwise comparisons within each dimension ($p_{\mathrm{adj}}<0.05$).
   }
    \label{fig:raw_corr}
\end{figure*}

We first examine whether the six main characters exhibit coordinated episode-to-episode variation in Power, Danger, and Structure. Figure~\ref{fig:raw_corr} shows the pairwise Pearson correlations calculated from the original episode-level time series after Bonferroni correction for the 15 character-pair comparisons within each dimension.

For Power, only three pairs show significant positive correlations: Monica--Rachel ($r=0.22$), Monica--Phoebe ($r=0.24$), and Monica--Chandler ($r=0.26$). We notice that all significant Power associations involve Monica, although the magnitudes of these correlations are modest. For Danger, only two pairs remain significant: Monica--Chandler ($r=0.20$) and Joey--Chandler ($r=0.28$). Structure exhibits a broader pattern of significant co-variation, with six pairs surviving the Bonferroni correction: Rachel--Joey ($r=0.22$), Monica--Joey ($r=0.24$), Phoebe--Chandler ($r=0.22$), Phoebe--Ross ($r=0.21$), Joey--Chandler ($r=0.21$), and Chandler--Ross ($r=0.20$). The generally weak correlations indicate that this common variation does not imply strong pairwise synchronization. Nevertheless, the associations that remain significant after Bonferroni correction reveal more localized patterns of temporal co-variation.

A consistent pattern is apparent across the three dimensions: all 45 pairwise correlation coefficients are positive. This systematic positive direction suggests that the characters' episode-level trajectories may contain a shared component, such that the characters tend to shift in the same direction.

We therefore further evaluate the measurement reliability of the episode-level trajectories and use partial correlations to determine whether any pairwise associations remain after accounting for variation shared with the other characters.

\subsubsection{Split-Half Reliability}
Before further interpreting these pairwise correlations, we assess the measurement reliability of the episode-level PDS trajectories using split-half reliability. Across the 18 character--dimension combinations, all correlations between the two randomly divided token halves are positive and statistically significant. After applying the Spearman--Brown correction, reliability estimates range from $0.330$ to $0.619$, indicating moderate stability in the episode-level measurements. The result is shown in Table~\ref{tab:split_half_reliability} in Appendix~\ref{Appendix:split_half}

Reliability varies across both dimensions and characters. Power shows the highest average Spearman--Brown reliability ($\bar{r}_{SB}=0.524$), compared with Structure ($\bar{r}_{SB}=0.445$) and Danger ($\bar{r}_{SB}=0.437$), suggesting that episode-to-episode variation along this dimension is measured more consistently than variation in Danger or Structure. The highest individual reliability is observed for Ross's Power trajectory ($r_{SB}=0.619$), followed by Rachel's Power ($r_{SB}=0.572$), Joey's Structure ($r_{SB}=0.567$), and Joey's Power ($r_{SB}=0.565$). The lowest estimates occur for Chandler's Danger ($r_{SB}=0.330$), Rachel's Structure ($r_{SB}=0.344$), and Phoebe's Power ($r_{SB}=0.377$). These results indicate that the episode-level PDS trajectories contain reproducible signal across independent dialogues. Consequently, weak pairwise correlations could be interpreted in the context of this moderate measurement reliability.

\subsubsection{Partial Correlations}

\begin{figure*}[t!]
    \centering
    \includegraphics[width=1.0\textwidth]{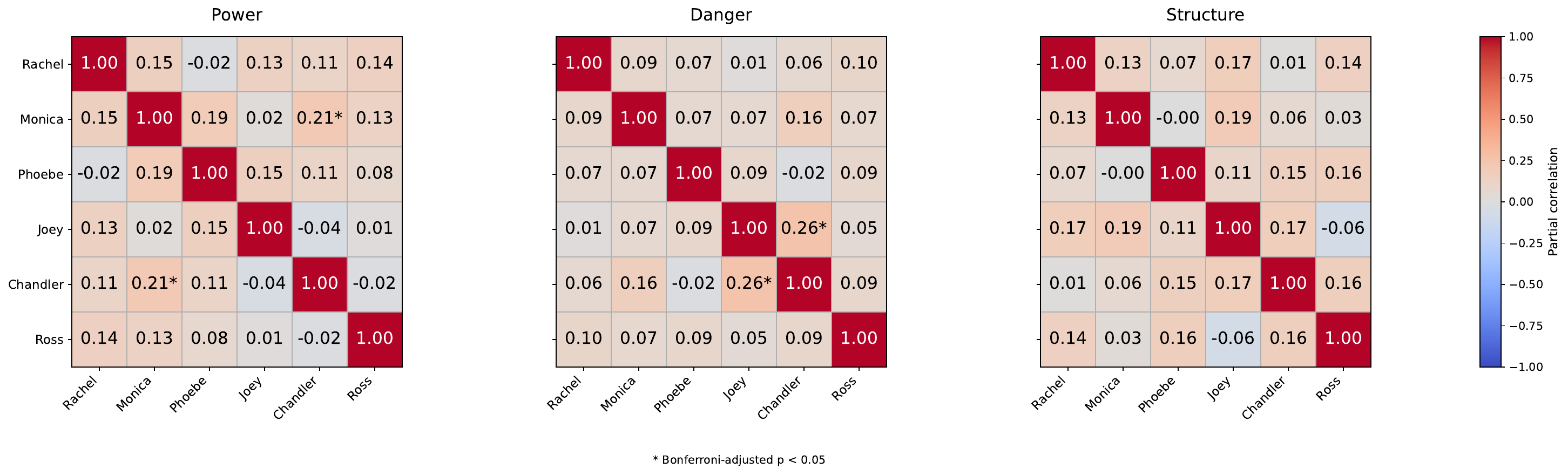}
    \caption{
    Partial correlations between the episode-level Power, Danger, and Structure trajectories of the six main characters. For each character pair, the partial correlation is computed from the original episode-level time series while controlling for the trajectories of the other four characters. Red indicates positive correlations, and blue indicates negative correlations. Positive values indicate that two characters tend to vary in the same direction after accounting for variation associated with the rest of the ensemble. Asterisks indicate correlations that remain statistically significant after Bonferroni correction for the 15 pairwise comparisons within each dimension ($p_{\mathrm{adj}}<0.05$).
    }
    \label{fig:partial_corr}
\end{figure*}

To determine whether the pairwise associations observed in the Pearson correlations reflect relationships specific to individual character pairs, we further compute partial correlations while controlling for the remaining four characters. As shown in Figure~\ref{fig:partial_corr}, the number of significant associations decreases substantially after accounting for the rest of the ensemble. For Power, only the Monica--Chandler positive association remains significant after Bonferroni correction ($r_{\mathrm{partial}}=0.21$). In comparison, the standard Pearson analysis identified three significant Power associations: Monica--Rachel, Monica--Phoebe, and Monica--Chandler. Thus, although Monica shows positive covariation with several characters, only her association with Chandler remains statistically detectable after the trajectories of the other four characters are controlled.

A similar reduction occurs for Danger. Of the two significant associations identified, only Joey--Chandler remains significant in the partial correlation analysis ($r_{\mathrm{partial}}=0.26$). For Structure, none of the 15 partial correlations survives Bonferroni correction. Thus, the partial correlation results substantially reduce the pairwise structure observed in the standard correlations.

Despite this, the Monica--Chandler relationship in Power and the Joey--Chandler relationship in Danger remain detectable and positive after controlling for the other four character trajectories, providing evidence of limited pair-specific conditional associations. The two surviving associations actually correspond to well-established relationships within the narrative. Monica and Chandler develop from friends into a long-term romantic partnership, while Joey and Chandler also maintain a close friendship for the whole series. They both repeatedly share scenes and narrative situations.

\section{Conclusions}

In this study, we apply the Power--Danger framework of ousiometrics to the complete ten-season dialogue corpus of \textit{Friends} and examine semantic dynamics at three levels: individual characters, character combinations, and pairwise trajectories, with Structure additionally considered in the pairwise analysis.

At the individual level, the six main characters are similar when their dialogue is aggregated across the entire series, occupying a close general Power--Safe region of the semantic space. The episode-level trajectories show substantial temporal variation, with multiple change points occurring across characters. Several of these transitions occur near identifiable narrative developments, suggesting that shifts in ousiometric trajectories can correspond to changes in narrative context.

At the combination level, Power varies more widely than Danger and appears more responsive to changes in group composition. However, the differences among character combinations do not provide evidence for clearly distinct combination types, and trajectory clustering similarly reveals little robust or well-separated structure. 

At the pairwise level, character trajectories generally move in the same direction across episodes, but these associations are modest and largely diminish after accounting for variation shared with the rest of the ensemble. Only Monica--Chandler in Power and Joey--Chandler in Danger retain detectable pair-specific conditional associations, while Structure shows no comparable pair-specific relationship. It suggests that the temporal co-variations of main characters are driven primarily by variation shared across the ensemble.

These findings extend the use of ousiometrics from describing aggregate patterns in language to examining semantic variation within a long-form ensemble narrative. By representing dialogue along interpretable semantic dimensions, the framework makes it possible to compare how language varies across characters, conversational groupings, and time within a common semantic space. In \textit{Friends}, this perspective suggests that Power and Danger are more informative as dimensions of changing narrative and interactional context. Ousiometrics therefore provides a representation for computational narrative analysis, connecting measurable variation in language to the evolving contexts.

\section{Limitations and Future Work}

This study has several limitations that also point to directions for future research. First, the analysis is based on a single television series. Several findings may therefore reflect properties specific to Friends or to the sitcom genre rather than general properties of narrative interaction. Comparative analyses across multiple series and genres are necessary to determine whether those patterns can generalize beyond Friends. Such comparisons could test, for example, whether the Power--Safe bias extends to other sitcoms and whether genres organized around conflict or threat, such as crime dramas, produce systematically different distributions in the Power--Danger space.

Second, the interaction analysis is restricted to the six main characters: Rachel, Monica, Phoebe, Joey, Chandler, and Ross. The 63 possible configurations and the subsequent combination-level analyses are therefore defined only with respect to this core ensemble. Scenes in Friends frequently include recurring and supporting characters, such as romantic partners, family members, and coworkers, whose presence may alter the interpersonal dynamics of the conversation. Consequently, two scenes classified under the same main-character combination may differ in their complete social configuration. Future work could incorporate supporting characters into the combination and examine whether the presence of third-party characters shifts the Power--Danger profile of interactions.

The third limitation is from the lexicon-based nature of ousiometrics itself. Each lexical item is assigned a relatively fixed score, and the semantic position of a dialogue or scene is estimated by aggregating these context-independent lexical values. This representation does not fully capture the contextual variability of meaning in dialogue. The same word may convey different levels of power or interpersonal danger depending on the pragmatic context. This is especially true for sitcoms, where sarcasm and irony are common and where the interactional meaning of an utterance may differ substantially from its literal content. The Power--Danger scores analyzed in this study should be interpreted just as lexicon-derived semantic signals. Future work could develop contextualized Power--Danger measures at the conversational level. Such an extension would help determine which interaction patterns are robust to contextual variation.

Finally, television narratives communicate meaning through multiple channels beyond words, including facial expression, gesture, and physical action. These signals may reinforce or contradict the semantic content of dialogue, particularly in a sitcom where irony and physical comedy are important devices. Extending the framework to the multimedia level would provide a more complete account of how interactional meaning is constructed in audio-visual narratives.



\section*{Acknowledgements}

The authors are grateful for
National Science Foundation Award \#2242829
(Science of Online Corpora, Knowledge, and Stories),
foundational support from MassMutual,
and
an anonymous philanthropic gift.



\clearpage

\addcontentsline{toc}{section}{References}

\bibliography{archetypes-of-LLMs}


\onecolumn

\appendix
\section{Appendices}

\setcounter{page}{1}
\renewcommand{\thepage}{A\arabic{page}}
\renewcommand{\thefigure}{A\arabic{figure}}
\renewcommand{\thetable}{A\arabic{table}}
\setcounter{figure}{0}
\setcounter{table}{0}

\renewcommand{\thesection}{A\arabic{section}}
\setcounter{section}{0}

\section{Character Combinations Remain}
\label{Appendix:combination_remain}

\begin{table}[ht!]
\centering
\begin{tabular}{lcc}
\hline
Character Combination & Scene Count & Word Count \\
\hline
Rachel, Monica, Phoebe, Joey, Chandler, Ross & 414 & 167,293 \\
Rachel, Ross & 177 & 35,331 \\
Monica, Chandler & 161 & 26,885 \\
Ross & 135 & 12,548 \\
Joey & 133 & 11,268 \\
Joey, Chandler & 130 & 20,790 \\
Phoebe & 128 & 14,275 \\
Rachel & 99 & 10,517 \\
Chandler & 91 & 7,322 \\
Joey, Chandler, Ross & 72 & 15,447 \\
Rachel, Joey & 71 & 14,815 \\
Rachel, Phoebe & 69 & 12,463 \\
Phoebe, Joey & 66 & 10,658 \\
Monica, Phoebe & 65 & 10,364 \\
Rachel, Monica, Phoebe & 64 & 14,021 \\
Monica & 63 & 4,763 \\
Monica, Joey, Chandler & 59 & 12,654 \\
Joey, Ross & 54 & 10,193 \\
Rachel, Monica & 53 & 8,585 \\
Rachel, Monica, Phoebe, Joey, Chandler & 48 & 14,741 \\
Monica, Phoebe, Chandler & 43 & 10,280 \\
Chandler, Ross & 42 & 6,609 \\
Monica, Phoebe, Joey, Chandler & 38 & 10,241 \\
Phoebe, Ross & 38 & 6,597 \\
Monica, Phoebe, Joey, Chandler, Ross & 35 & 8,276 \\
Rachel, Monica, Phoebe, Ross & 33 & 8,974 \\
Rachel, Monica, Joey, Chandler, Ross & 33 & 9,980 \\
Rachel, Phoebe, Ross & 30 & 6,767 \\
Monica, Joey & 29 & 4,248 \\
Monica, Ross & 27 & 3,487 \\
Monica, Chandler, Ross & 27 & 5,881 \\
Rachel, Monica, Chandler, Ross & 26 & 6,062 \\
Rachel, Phoebe, Joey, Ross & 25 & 7,622 \\
Rachel, Monica, Phoebe, Chandler, Ross & 24 & 8,939 \\
Rachel, Monica, Ross & 24 & 5,003 \\
Rachel, Monica, Joey, Chandler & 24 & 6,128 \\
Rachel, Joey, Ross & 24 & 6,751 \\
Rachel, Joey, Chandler, Ross & 23 & 5,865 \\
Rachel, Phoebe, Joey & 23 & 4,864 \\
Rachel, Monica, Phoebe, Joey, Ross & 23 & 4,658 \\
Rachel, Chandler & 22 & 4,080 \\
Monica, Joey, Chandler, Ross & 21 & 5,570 \\
Phoebe, Chandler & 21 & 3,046 \\
Rachel, Monica, Phoebe, Chandler & 19 & 4,905 \\
Rachel, Monica, Chandler & 19 & 4,819 \\
Rachel, Joey, Chandler & 18 & 4,292 \\
Phoebe, Joey, Chandler, Ross & 18 & 4,674 \\
Phoebe, Joey, Chandler & 17 & 3,369 \\
Monica, Phoebe, Chandler, Ross & 17 & 4,587 \\
Rachel, Phoebe, Joey, Chandler, Ross & 17 & 4,716 \\
Phoebe, Joey, Ross & 16 & 3,360 \\
Rachel, Monica, Phoebe, Joey & 15 & 4,444 \\
Monica, Phoebe, Ross & 14 & 2,936 \\
Rachel, Chandler, Ross & 14 & 3,761 \\
Monica, Phoebe, Joey & 13 & 2,455 \\
Rachel, Phoebe, Chandler, Ross & 12 & 2,252 \\
\hline
\end{tabular}
\caption{ Combinations remain for combination-level analysis.}
\label{table:combination_remain}
\end{table}

\begin{table}[h]
\centering
\begin{tabular}{lcc}
\hline
Character Combination & Scene Count & Word Count \\
\hline
Rachel, Monica, Joey & 11 & 1962 \\
Rachel, Monica, Joey, Ross & 9 & 1963 \\
Rachel, Phoebe, Chandler & 8 & 2183 \\
Phoebe, Chandler, Ross & 7 & 1435 \\
Rachel, Phoebe, Joey, Chandler & 6 & 1130 \\
Monica, Joey, Ross & 5 & 731 \\
Monica, Phoebe, Joey, Ross & 2 & 427 \\
\hline
\end{tabular}
\caption{ Combinations that have been removed for further analysis.}
\label{table:combination_filter}
\end{table}

\clearpage

\section{Change Point of Each Character}
\label{Appendix:change_point}

\begin{table}[ht!]
\centering
\small
\begin{tabular}{llllll}
\hline
\textbf{Dimension} & \textbf{Character} & \textbf{\#} & \textbf{Position} \\
\hline

Power & Rachel & CP1  & S02E23 \\
      &        & CP2  & S03E09 \\
      &        & CP3  & S03E19 \\
      &        & CP4  & S04E05 \\
      &        & CP5  & S05E19 \\
      &        & CP6  & S06E13 \\
      &        & CP7  & S07E07 \\
      &        & CP8  & S07E17 \\
      &        & CP9  & S08E04 \\
      &        & CP10 & S08E17 \\
\addlinespace

Power & Monica & CP1 & S06E22 \\
      &        & CP2 & S07E08 \\
      &        & CP3 & S08E23 \\
      &        & CP4 & S09E23 \\
\addlinespace

Power & Phoebe & CP1 & S03E02 \\
      &        & CP2 & S04E06 \\
      &        & CP3 & S05E03 \\
      &        & CP4 & S05E22 \\
      &        & CP5 & S06E19 \\
      &        & CP6 & S08E13 \\
      &        & CP7 & S08E23 \\
      &        & CP8 & S09E10 \\
\addlinespace

Power & Joey & CP1 & S01E16 \\
      &      & CP2 & S03E02 \\
      &      & CP3 & S03E18 \\
      &      & CP4 & S05E19 \\
      &      & CP5 & S06E14 \\
      &      & CP6 & S07E15 \\
      &      & CP7 & S08E22 \\
      &      & CP8 & S09E10 \\
      &      & CP9 & S09E21 \\
\addlinespace

Power & Chandler & CP1 & S01E24 \\
      &          & CP2 & S04E10 \\
      &          & CP3 & S05E12 \\
      &          & CP4 & S06E04 \\
      &          & CP5 & S06E18 \\
      &          & CP6 & S07E03 \\
      &          & CP7 & S07E14 \\
      &          & CP8 & S08E05 \\
      &          & CP9 & S08E21 \\
\addlinespace

Power & Ross & CP1  & S02E11 \\
      &      & CP2  & S02E22 \\
      &      & CP3  & S03E23 \\
      &      & CP4  & S04E12 \\
      &      & CP5  & S05E09 \\
      &      & CP6  & S05E22 \\
      &      & CP7  & S06E12 \\
      &      & CP8  & S07E16 \\
      &      & CP9  & S08E09 \\
      &      & CP10 & S10E03 \\

\hline
\end{tabular}
\caption{PELT-detected change points for the Power dimensions of the six main characters in \textit{Friends}.}
\label{Tab:change_point_power}
\end{table}

\clearpage

\begin{table}[ht!]
\centering
\small
\begin{tabular}{llllll}
\hline
\textbf{Dimension} & \textbf{Character} & \textbf{\#} & \textbf{Position} \\
\hline

Danger & Rachel & CP1 & S01E11 \\
       &        & CP2 & S01E22 \\
       &        & CP3 & S03E08 \\
       &        & CP4 & S04E11 \\
       &        & CP5 & S05E14 \\
       &        & CP6 & S06E19 \\
\addlinespace

Danger & Monica & CP1 & S03E17 \\
       &        & CP2 & S04E06 \\
       &        & CP3 & S07E02 \\
       &        & CP4 & S08E04 \\
       &        & CP5 & S09E15 \\
       &        & CP6 & S10E01 \\
\addlinespace

Danger & Phoebe & CP1 & S01E24 \\
       &        & CP2 & S02E13 \\
       &        & CP3 & S04E24 \\
       &        & CP4 & S06E08 \\
\addlinespace

Danger & Joey & CP1 & S01E11 \\
       &      & CP2 & S02E10 \\
       &      & CP3 & S05E10 \\
       &      & CP4 & S08E17 \\
       &      & CP5 & S09E03 \\
\addlinespace

Danger & Chandler & CP1 & S01E12 \\
       &          & CP2 & S01E23 \\
       &          & CP3 & S02E09 \\
       &          & CP4 & S03E07 \\
       &          & CP5 & S04E24 \\
       &          & CP6 & S05E17 \\
       &          & CP7 & S06E06 \\
       &          & CP8 & S07E18 \\
\addlinespace

Danger & Ross & CP1 & S01E21 \\
       &      & CP2 & S03E02 \\
       &      & CP3 & S04E02 \\
       &      & CP4 & S05E01 \\
       &      & CP5 & S05E20 \\
       &      & CP6 & S06E07 \\
       &      & CP7 & S06E17 \\
       &      & CP8 & S07E03 \\

\hline
\end{tabular}
\caption{PELT-detected change points for the Danger dimensions of the six main characters in \textit{Friends}.}
\label{Tab:change_point_danger}
\end{table}

\clearpage

\section{Character Combinations Retained for Trajectory Clustering}
\label{Appendix:Character_Combinations}

\begin{table}[h!]
\centering
\begin{tabular}{l c c}
\hline
\textbf{Combination}  & \textbf{Valid Seasons} & \textbf{Total Scenes} \\
\hline
Chandler  & 8 & 88 \\
Monica + Chandler  & 8 & 159 \\
Rachel + Monica + Phoebe + Joey + Chandler  & 8 & 44 \\
Rachel + Monica + Phoebe + Joey + Chandler + Ross  & 10 & 414 \\
Monica + Phoebe  & 9 & 63 \\
Phoebe + Joey  & 8 & 63 \\
Rachel  & 10 & 99 \\
Rachel + Monica + Phoebe  & 9 & 63 \\
Rachel + Phoebe  & 10 & 69 \\
Joey + Chandler  & 9 & 128 \\
Rachel + Ross  & 10 & 177 \\
Ross & 10 & 135 \\
Joey + Chandler + Ross & 9 & 71 \\
Phoebe  & 10 & 128 \\
Joey  & 10 & 133 \\
\hline
\end{tabular}
\caption{Character Combinations Retained for Trajectory Clustering}
\label{table:cluster_combination}
\end{table}

\section{Split-half Reliability}
\label{Appendix:split_half}

\begin{table}[htbp]
\centering
\caption{Split-half reliability of episode-level Power, Danger, and Structure scores for the six main characters. For each character and dimension, Pearson's $r$ measures the correlation between scores calculated from two random halves of the dialogue tokens across 236 episodes. The final column reports the Spearman--Brown corrected reliability coefficient.}
\begin{tabular}{llrrr}
\toprule
Character & Dimension & $r_{\mathrm{half}}$ & $p$ & $r_{\mathrm{SB}}$ \\
\midrule

Rachel
    & Power     & 0.400 & $<.001$ & 0.572 \\
    & Danger    & 0.257 & $<.001$ & 0.409 \\
    & Structure & 0.207 & 0.001    & 0.344 \\
\addlinespace

Monica
    & Power     & 0.341 & $<.001$ & 0.509 \\
    & Danger    & 0.307 & $<.001$ & 0.470 \\
    & Structure & 0.245 & $<.001$ & 0.394 \\
\addlinespace

Phoebe
    & Power     & 0.232 & $<.001$ & 0.377 \\
    & Danger    & 0.263 & $<.001$ & 0.416 \\
    & Structure & 0.255 & $<.001$ & 0.406 \\
\addlinespace

Joey
    & Power     & 0.394 & $<.001$ & 0.565 \\
    & Danger    & 0.369 & $<.001$ & 0.539 \\
    & Structure & 0.395 & $<.001$ & 0.567 \\
\addlinespace

Chandler
    & Power     & 0.337 & $<.001$ & 0.504 \\
    & Danger    & 0.198 & 0.002    & 0.330 \\
    & Structure & 0.364 & $<.001$ & 0.534 \\
\addlinespace

Ross
    & Power     & 0.449 & $<.001$ & 0.619 \\
    & Danger    & 0.298 & $<.001$ & 0.459 \\
    & Structure & 0.272 & $<.001$ & 0.428 \\

\bottomrule
\end{tabular}
\label{tab:split_half_reliability}
\end{table}

\end{document}